\documentclass{aa}  

\usepackage{graphicx}
\usepackage{txfonts}
\usepackage{lipsum}
\usepackage{subcaption}         
\usepackage{lscape}             
\usepackage{placeins}           

\usepackage[T1]{fontenc}
\usepackage{lmodern}
\usepackage{amssymb}
\usepackage{amsmath}

\usepackage{float}
\usepackage{hyperref}
\usepackage{xurl}
\usepackage{multirow}
\usepackage{microtype}
\usepackage{natbib}
                                
\begin{document}

\title{(An)Isotropy in Pantheon+ and Type Ia supernova samples:
    intrinsic limits of directional tests. }


%
%
%

   \author{A. Quintana-Estell\'es\inst{1,2}\corrauth{antonio.quintana@iff.csic.es}   
        \and P. Ruiz-Lapuente\inst{1,2}\email{pilar@icc.ub.edu}
        }

   \institute{Instituto de F\'isica Fundamental, Consejo Superior de Investigaciones Cient\'{\i}ficas, c/. Serrano 121,
     E-28006, Madrid, Spain
   \and Institut de Ci\`encies del Cosmos (UB-IEEC), c/. Mart\'{\i} i Franqu\'es 1, E-08028, Barcelona, Spain}

   \date{Received May , 2026}

 
  \abstract
      {The use of methods that investigate the value of the
        Hubble constant H$_{0}$ in different patches (60$^{\circ}$ or 90$^{\circ}$ size)
        across the sky to probe the statistical isotropy of the Universe  using
        large SNe Ia databases has led to contradictory claims of either anisotropy or
        isotropy. The anisotropy directions vary amongst research works. }
      {The objective of this paper is to clarify the abovementioned claims and
        study the lack of basis for depicting directions of anisotropy with
        the present SNe Ia samples. We explain the type of limitation
        embedded in the
        SN Ia lightcurve method to determine the isotropy of H$_{0}$  and the corresponding
         consequences. } 
      {The widely used analysis through the Region Fitting and the
        Hemisphere Comparison methods is done here using the Pantheon+ database,  
        simulating 2000
        distinct directions in the sky within a Bayesian Markov
        Chain Monte Carlo
        approach. We also study a smaller SNe Ia database, the Carnegie
        Supernova Project sample, leading to
        a similar kind of result as that from the Pantheon+ sample.
        We investigate the validity of the directions found for anisotropy within
        these analyses.}
      {We have found that within the tests used here, the Region
        Fitting method and the Hemisphere Comparison method, one
        can not determine with
        robustness the direction of an anisotropy of H$_{0}$ using the
        present SNe Ia large data samples.
        This is intrinsic to the way H$_{0}$ is obtained with the SNe Ia
        lightcurve method. }
      { Achieving robust constraints will require a quite uniform sky coverage from
        larger SNe Ia samples with improved systematics.}


   \titlerunning{Directional Tests of Isotropy}

   \maketitle

\section{Introduction}

\noindent
Type Ia supernovae (SNe Ia) at high $z$ led to the discovery of the acceleration
of the Universe (Riess et al. 1998; Perlmutter et al. 1999) and dark energy.
Large compilations of SNe Ia have recently suggested a tentative evolution
in time of such unknown major component of the energy density of the Universe
(Rubin et al. 2025; DESyr5, DESI\footnote{DESyr5 stands for
''Dark Energy Survey year 5'', DESI
for ''Dark Energy Spectroscopic Instrument'', SH0ES for
''Supernova H0 for the Equation of State of Dark Energy''.}).

\bigskip

\noindent
Apart from this use in cosmology, a puzzle emerged some  years ago and still
requires an explanation: the expansion rate
of the Universe today, as obtained from
SNe Ia calibrated with Cepheids,  gives a value of H$_{0}$ higher than that
measured
from the Cosmic Microwave Background (CMB).
The value of H$_{0}$ derived from the CMB is
  67.4$\pm$0.5 km s$^{-1}$ Mpc$^{-1}$ (Planck Collaboration 2020) and the 
 {\it SH0ES}
 value of  H$_{0}$ =73.3$\pm$0.9 km s$^{-1}$ Mpc$^{-1}$
 (Murakami et al et. 2023) had a discrepancy at the 5.7$\sigma$ level
 or even  at the 6$\sigma$ level (Riess et al. 2025).
 The most recent results from polarization of the CMB obtained by
 the Third-Generation camera for the South Pole Telecope (SPT-3G)
 (Camphuis et al. 2026)
 report a Hubble constant of H$_{0}$ = 66.66 $\pm$ 0.60 km s$^{-1}$ Mpc$^{-1}$,
 6.2$\sigma$ 
 away from local measurements from {\it SH0ES}.

 \bigskip

 \noindent
 A question that already arose when debating on the acceleration of
 the expansion of the Universe
(Riess et al. 1998; Perlmutter et al. 1999) 
 was whether the FLRW metric, which is valid for 
an homogeneous and isotropic Universe, was a legitimate assumption. 
 If that was not the case, the findings from cosmological SNe Ia 
 would be compromised. This had prompted interest in testing whether the
 Universe is isotropic and homogeneous on large scales ($>$ 260 h$^{-1}$ Mpc)
  matter being evenly distributed and with no preferred directions. That would
 fulfill the Cosmological Principle
 (CP) and justify the assumption of the $\Lambda$CDM metric. For this, 
 bulk flows arising from
 matter structure distribution which generates peculiar velocity fields should
 not exceed the $\Lambda$CDM expectations  on scales of $>$ 260 h$^{-1}$.
 There are plenty of research works addressing
 whether this is the case. We refer here to Di Valentino et al. (2025),
 and references
 therein, for a complete update and discussions on this topic. 
 We also refer to the review on large-scale peculiar velocities
 in the universe by  Tsagas, Perivolaropoulos and Asvesta (2026). 
 Recently, Lopes et al.(2024) have
 determined, using Pantheon+ SNe Ia, a bulk flow motion towards a direction
 close to the Shapley supercluster of 132.14 $\pm$ 109.3 km s$^{-1}$ in velocity
 at the
 effective distance of 102.83 $\pm$ 10.2 Mpc.  Such motion would be still
 consistent with the $\Lambda$CDM model. However, larger motions
 have  been claimed at other effective distances by Watkins et al. (2023),
 Whitford et al.(2023) using the Cosmic-Flows4 (Tully et al. 2023).
 The samples used in those three research works of the Cosmic-Flows4 project
 are much larger than those from SNe Ia. Though, at present systematics
 seems to be reevaluated.

 \bigskip

 \noindent
  The  debate both on the isotropy of the Hubble flow and
 on the compatibility of cosmic flows  uses
 a variety of methods. A large number of tests use distance indicators
 to examine this question, such as the Tully-Fisher (T-F) relation, the
 Fundamental Plane (F-P) relation,  the Surface Brightness Fluctuations
 (SBF), Type Ia supernovae (SNe Ia) and Type II SNe (SNe II).
 We will examine  the part of Type Ia SNe, often used on their own.
 
\bigskip 

\noindent
The use of approaches that investigate the value of the
Hubble constant H$_{0}$
or the deceleration parameter q$_{0}$ in different patches
(60$^{\circ}$ or 90$^{\circ}$ size) across the sky to probe the statistical
isotropy of the Universe has led in some cases to claims of
anisotropy and isotropy. There has not been consesus on this matter
using different cosmological SNe Ia databases.  
Many authors have used the Pantheon+ database  (Scolnic et al. 2022;
Brout et al. 2022) to perform these tests
(Bengaly, Alcaniz \& Pigozzo 2024; Mc Conville \& Colg\'ain 2023). 
Earlier works had used smaller samples such as the Pantheon (Scolnic et al.
2018) in Zhao, Zhou \& Chang (2019); 
the Constitution sample
(Hicken et  al. 2009) in Kalus et al. (2013); the Union 2 sample
(Amanullah et al. 2010) in  Antoniou \& Perivolaropoulos (2010) and in 
Colin et al. (2011); the Joint Light-Curve Analysis (JLA) (Betoule et al. 2014)
in Deng \& Wei  (2018). Independently of the results, that could
depend on the amount of SNe Ia in the database, there is the question of
the nature of the scatter in those samples which has a very important 
effect in the validity of these approaches.

\bigskip

\noindent
We start from the real uncertainty of the individual H$_{0}$
measurements in Pantheon+ (Brout et al. 2022; Scolnic et al. 2022) samples.
The point to point variation of H$_{0}$  in the Pantheon+ sample and
in other SNe Ia  samples is basically of the order of 8--9 km s$^{-1}$.
Part of this error is due to an intrinsic limitation of the SNe Ia light curve
method to determine distances, which is  of $\sigma_{int}$ $=$ 0.11 mag (see
for instance Mandel, Narayan \& Kirshner 2011). This limitation in the method
accounts for a 4 km s$^{-1}$ Mpc$^{-1}$\footnote{An error of $\sigma_{\mu}$ = 0.11 mag would result, for a value of
$H_{0}$ = 73 km s$^{-1}$ Mpc$^{-1}$, very approximately in an error in H$_{0}$
of
3.7 km s$^{-1}$ Mpc$^{-1}$, irrespective of the distance.
This limit is calculated from the following:

\noindent

\noindent
$ \sigma_{H_{0}} = H_{0} \sqrt{\left(\frac{\sigma_{z}}{z}\right)^{2} +
  \left(\frac{\sigma_{D}}{D}\right)^{2}}$.

\noindent
If we dismiss, for being close to 0,
$\frac{\sigma_{z}}{z}$, then
 $\sigma_{H_{0}} \sim H_{0} \left(\frac{\sigma_{D}}{D}\right)$

\noindent
and the error in D is :

$\sigma_{D} = \frac{1}{5} {\rm ln(10)} \sigma_{\mu} D$. This gives: 

$\sigma_{H_{0}} \sim \frac{1}{5}\,\ln(10)\,\sigma_{\mu}\, H_{0}$
}.
  But, in fact, the larger variation, of the
order of 8--9 km s$^{-1}$ Mpc$^{-1}$, is due to other errors such as the anchor in Cepheids
(where in fact there could be an anisotropy in the 
absolute calibration  of the period--luminosity relation
of the  Cepheids, given their low number of calibrators)
or other factors such as the effect of existing 
peculiar velocities affecting  SNe Ia. 

\bigskip

\noindent
We show here the effect that such H$_{0}$ point to
point variation has, even considering only the intrinsic error of the light curve method, in the analysis of the anisotropy of H$_{0}$ in the sky
and in the direction of such anisotropy. We do this with the Pantheon+
sample and compare it with research that has not taken this effect into account. We show  the effect, as well,  in  a smaller but
carefully treated  SNe Ia sample 
as that of the  Carnegie Supernova Project (Freedman et al.
2019). 

\bigskip

\noindent
Thus, we study the 
pattern in the measurement of H$_{0}$ across the sky and test
the ability to point to anisotropies using the SNe Ia samples.  
We examine the anisotropy/isotropy claims done so far to shed
light on this important question. 

\bigskip

\noindent
This paper is structured as follows. In section 2, we present 
a comprehensive overview of the Pantheon+  sample, illustrating its distribution in 
Galactic coordinates and in redshift. This section 
introduces the framework for infering cosmological parameters from the luminosity data of 
SNe Ia. We describe in detail our methodology in section 3. 
The main results are presented in sections 4  and  5. 
In section 6 we provide the corresponding discussion, and we conclude with our 
findings in section 7.

\bigskip

\section{Pantheon+ dataset}

\noindent
We explore the Pantheon+ sample (Brout et al. 2022; Scolnic et al. 2022), which includes 1701 light curves 
of 1550 distinct SNe Ia from 18 different surveys, confirmed through spectroscopy and with redshifts 
ranging $0.001 < z < 2.26$. This sample can be used as it is or in combination with SH0ES results 
(Riess et al. 2022). When combined, the inclusion of 42 Cepheids distance measurements, in the same 
host galaxies as some supernovae, enables the calibration of the absolute magnitude of \mbox{SNe Ia} 
using the cosmic distance ladder method. We show in Figure 1
the SNe Ia distribution in Galactic coordinates and that in redshift in Figure 2.

\begin{figure}[htbp]
\centering
\includegraphics[width=.46\textwidth]{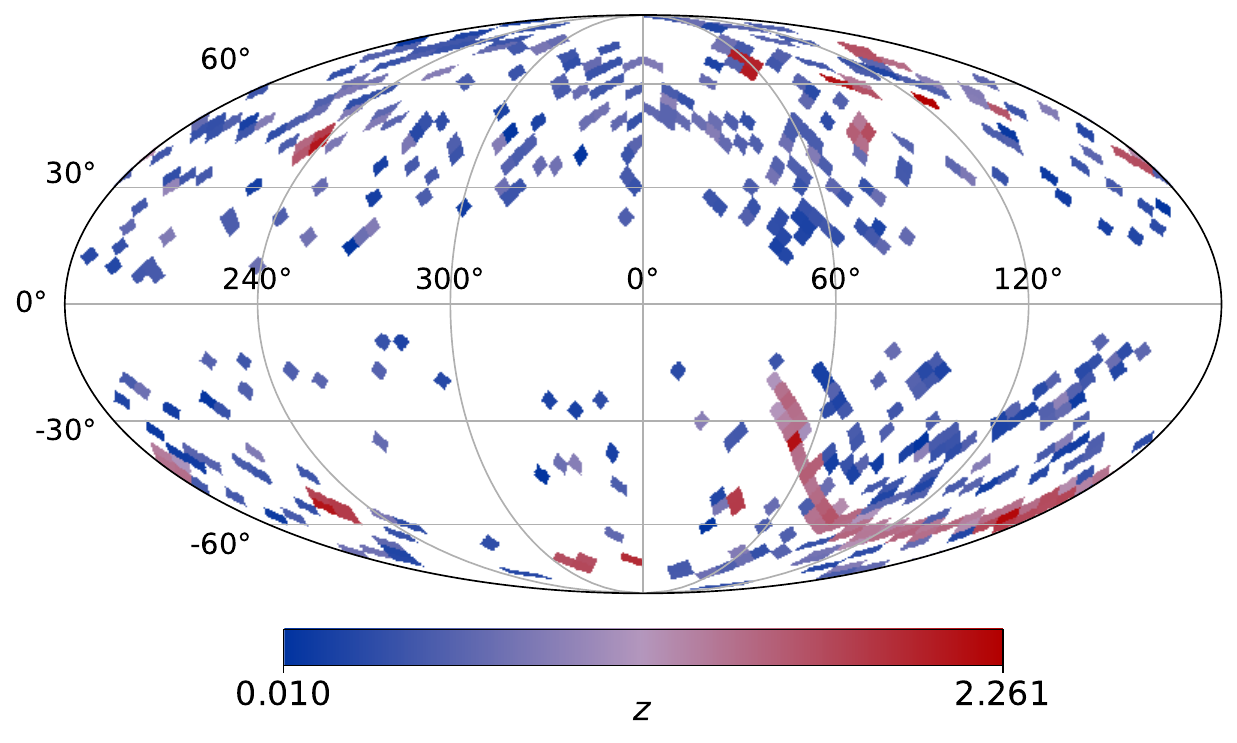}\\
\caption{SNe Ia distribution in galactic coordinates, with the color bar representing their redshift.}
\label{Figure 1}
\end{figure}

\begin{figure}[htbp]
\centering
\includegraphics[width=0.4\textwidth]{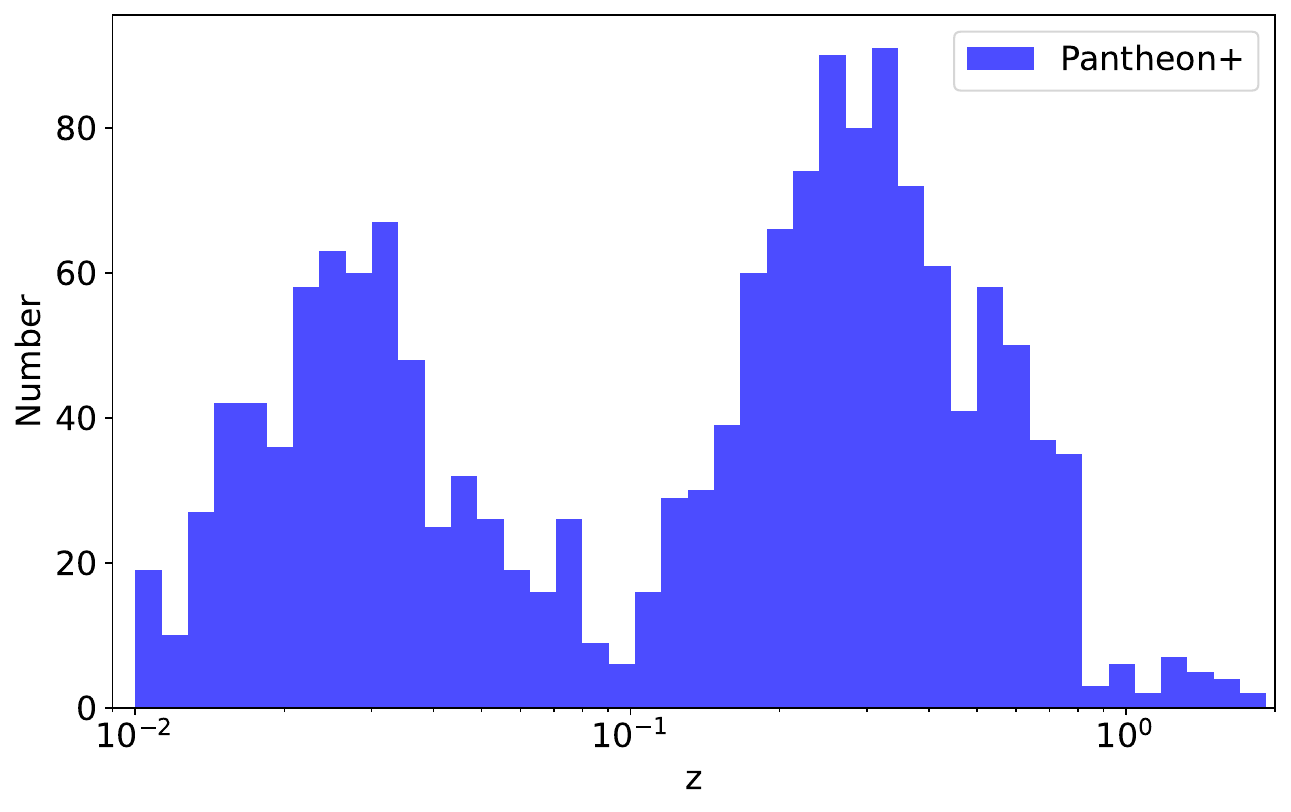}
\caption{Redshift histogram for Pantheon+  dataset.}
\label{Figure 2}
\end{figure}

\bigskip

\bigskip

$\chi(z_\text{cmb})$ is the comoving distance, defined as:

\begin{equation}
\label{eq:chi}
\chi\left(z_\text{cmb}\right)=\int_0^{z_\text{cmb}} \frac{dz}{H(z)}
\end{equation}

\begin{figure}[htbp]
\centering
\includegraphics[width=.46\textwidth]{"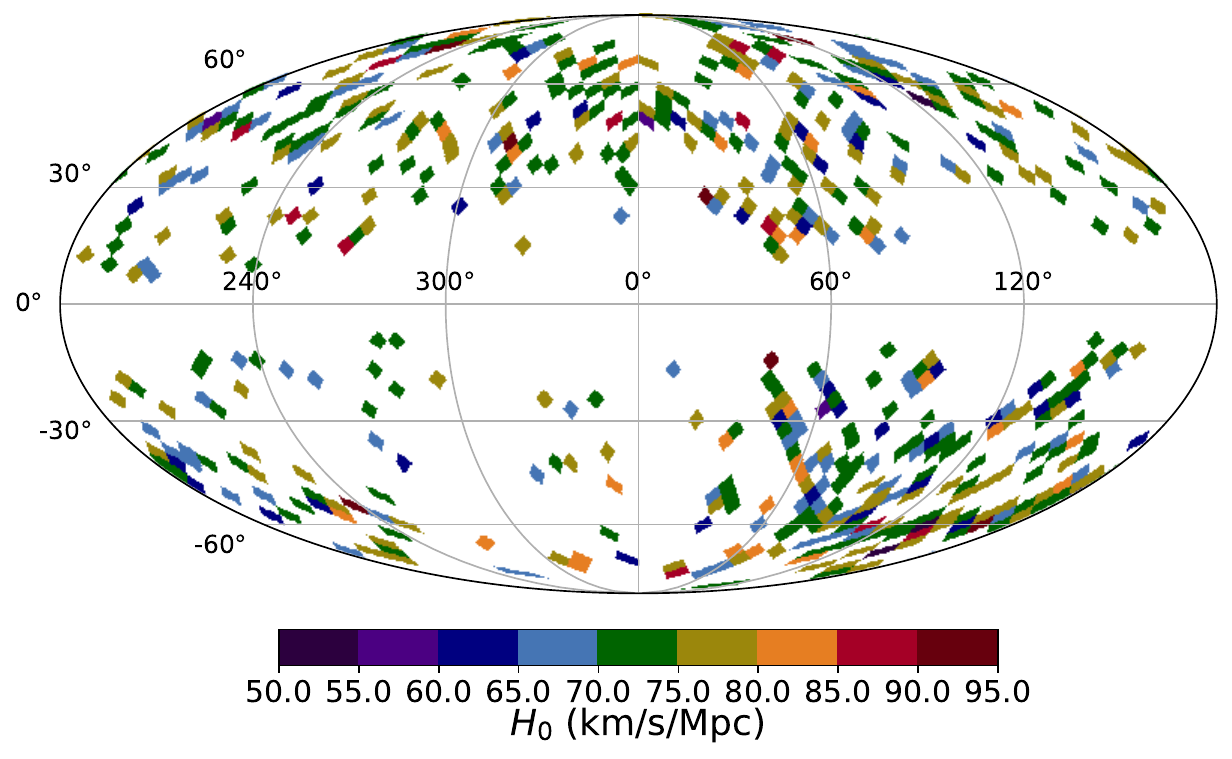"}\\
\caption{Individual $H_0$ values derived by inverting the distance–redshift
relation for SNe Ia. Pantheon+ dataset.}
\label{Figure 3}
\end{figure}

\begin{figure}[htbp]
\centering
\includegraphics[width=.4\textwidth]{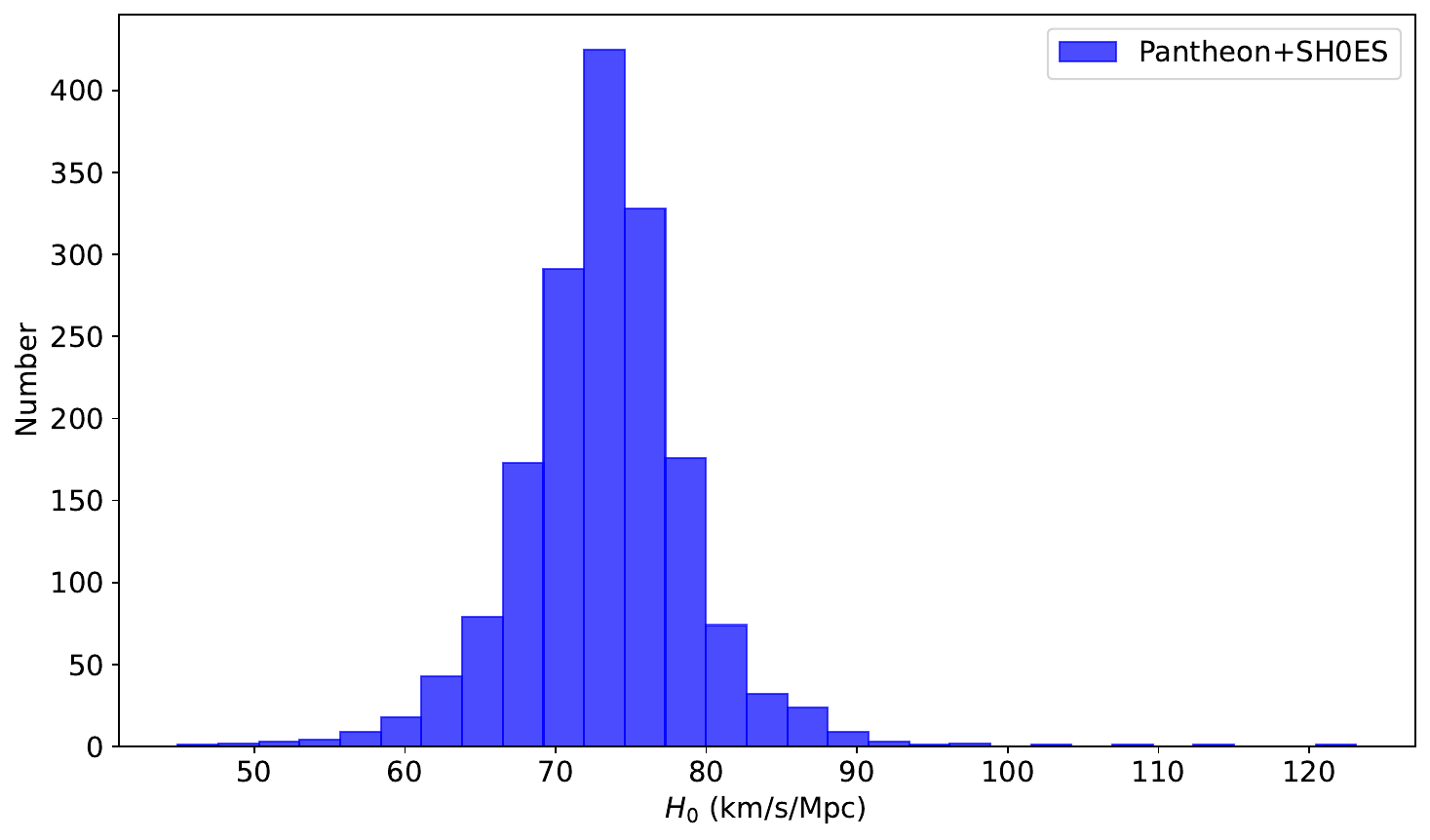}\\
\caption{ Histogram of individual $H_0,i$ values.}
\label{Figure 4}
\end{figure}

\noindent
More interestingly, we present in Figure 3 the Hubble constant ($H_0$) values computed 
for individual SNe Ia using Pantheon+ , by inverting the distance–redshift relation 
using the original dataset values. The luminosity distance is derived from the distance 
modulus, and $H_0$ is calculated assuming a second-order cosmographic expansion in redshift, 
with fixed deceleration and jerk parameters, $q_0 = -0.55$ and $j_0 = 1.0$. This approach 
yields a preliminary local estimate of $H_0$ for each SN Ia ($H_{0,i}$),
enabling visualization of its 
directional distribution. We observe that the resulting $H_0$ values, H$_{0,i}$ appear to be distributed 
in a seemingly random manner, with no apparent trend or structure. Moreover, the individual $H_{0,i}$ 
values exhibit a considerable spread, ranging from 50 to almost 100~km~s$^{-1}$~Mpc$^{-1}$ (see Figure 4).

\bigskip

\noindent
We start by introducing the fundamental relationship between the luminosity distance, $D_\text{L}$, 
and the expansion history, $H(z)$. For a spatially flat universe, this relationship is expressed by 
equation 2:
\begin{equation}
D_\text{L}(z_\text{hel}, z_\text{cmb}) = (1+z_\text{hel})\frac{c}{H_0}\chi(z_\text{cmb})
\end{equation}

\noindent
where $c$ is the speed of light; $H_0$ is the Hubble constant; $z_\text{hel}$ is the heliocentric 
redshift; $z_\text{cmb}$ is the redshift in the CMB frame due to the expansion of the 
universe, 
with $H(z)$ depending on the underlying cosmological model. For a Flat-$\Lambda$CDM model, it takes 
the form:

\begin{equation}
\begin{aligned}
H(z) &= H_0\sqrt{\Omega_m (1+z)^3 + (1 - \Omega_m)} \\
\end{aligned}
\end{equation}

\noindent
where $\Omega_{m}$ is the matter density parameter. The observed distance modulus, 
$\mu_\text{obs,i}=m_i-M$, for each SN Ia is defined in terms of its apparent magnitude, 
$m_i$, and absolute magnitude, $M$, which is degenerate with $H_0$. This degeneracy can 
be removed by using distance measurements to Cepheids located in the same host 
galaxies as the SNe Ia, as provided by SH0ES  (Riess et al. 2022).
For Pantheon+ 
the observed distance moduli are estimated using equation (1) from Brout et al. (2022).\footnote{

\noindent
For Pantheon+, the redshift used in equation 1 is the so-called 
\textit{Hubble Diagram redshift} as defined in equation (7) of Carr et al.
(2022), which accounts 
for both our motion relative to the CMB and the peculiar velocity of the source.
The studies of Carr et al. (2022) are built upon the idea of 
replacing individual host galaxy redshifts with average redshifts of host galaxy groups 
(Peterson et al. 2022). Such redshift is: 

\begin{equation}
z_{\rm HD} = \frac{1 + z_{\rm CMB}}{1 + z_{\rm p}} - 1\end{equation}

\noindent
The \textit{Hubble diagram redshift} $z_{\rm HD}$ requires knowledge of the SN
host’s peculiar redshift z$_{p}$.  The  z$_{\rm CMB}$ is the CMB redshift and
z$_{\rm p}$ is the redshift corresponding to the peculiar velocity,
which is modeled from peculiar velocity groups reconstruction by Peterson et al. (2022).

}

\bigskip

\noindent
Pantheon+ has been crucial in the development 
of tests of cosmic isotropy and in attempts to describe cosmic bulk motions.  
However, the limitation in the method of 4 km s$^{-1}$ Mpc$^{-1}$  in H$_{0,i}$
confines as well the
possibility to determine 
bulkflows. It is possible to determine them at low z, but as z grows
the constraint from H$_{0,i}$ needs a larger bulk flow to be detected.
 This can be seen from the relation between
peculiar velocities, luminosity distance D$_{L}$  and H$_{0}$:

\begin{equation}
\sigma_{v_{pec}} = \frac{1}{\sqrt{2}} \sqrt{D_{L}^{2}\sigma_{H_{0}}^{2} +
                     H_{0}^{2}\sigma_{D_{L}}^{2}}
\end{equation}

\noindent
The value  of the Hubble flow in the sky, with the distribution of individual H$_{o,i}$,
is a source for studing bulkflows at a very low z, but it is not so for larger z. 
The overall distribution of matter and the produced field of peculiar velocities leave
imprints on H$_{0}$ in all directions of the sky.  

\bigskip

\noindent
Here the full Pantheon+SH0ES dataset will be analyzed using the Region Fitting (RF) 
method (Hu et al. 2023) and the Hemisphere Comparison (HC) method
(Schwarz and Weinhorst 2007). The RF method
incorporates the full covariance matrix to account for both statistical 
and systematic uncertainties. For this dataset, we investigate the impact of the Cepheid calibrator 
subsample. That is, for any given region of the sky, we examine the effect of using either the full 
set of 42 Cepheid calibrators or only those located within the region under study. We additionally 
test whether the directions of maximum anisotropy are consistent with statistical noise and sample 
variance. Thus, given that the SNe Ia lightcurve method has
an intrinsic error of residuals of  $\sigma_{int} \simeq 0.11$ mag, that
corresponds to variations of approximately 
$4~\mathrm{km~s^{-1}~Mpc^{-1}}$ in H$_{0}$,
we perform simulations to test whether the directions of maximum 
anisotropy remain stable under such variations in the distance modulus. We also apply  the HC method  to Pantheon+
and similarly explore the effect of the distance uncertainties on the results.


\section{Methodology}

\noindent
Therefore, in order to compare observations 
with theory, the theoretical distance modulus for each SNe Ia, given its redshift, $z_i$, 
and a set of cosmological parameters, $\Theta$, is:

\begin{equation}
\mu_\text{th}(z_i,\Theta)=5\log_{10}\left(\frac{D_\text{L}(z_i,\Theta)}{\text{Mpc}}\right)+25
\end{equation}

\noindent
For Pantheon+, the best-fit of the cosmological parameters can be obtained by minimizing a 
$\chi^2$ likelihood, $\mathcal{L}$, defined as:

\begin{equation}
-2\ln{\left(\mathcal{L}\right)} = \chi^2 = \Delta\mu_i C_{ij}^{-1}\Delta\mu_j^T
\end{equation}

\noindent
where $\Delta\mu_i=\mu_\text{obs,i}-\mu_\text{th}(z_i,\Theta)$, and $C_{ij}^{-1}$ is the 
inverse of the covariance matrix, which accounts for both systematic and statistical 
uncertainties, as well as the expected correlations among the \mbox{SNe Ia} within the 
sample when evaluating cosmological models. Each element of the covariance matrix represents 
the covariance between pairs of SNe Ia in the sample (Vincenzi et al.
2024). 

\bigskip

\noindent
The minimization of the $\chi^2$ likelihood function is made using a Bayesian Markov Chain 
Monte Carlo (MCMC) approach. Among the tools used in this analysis, we employ the CosmoSIS 
framework (Zuntz 2015), a modular tool designed for cosmological parameter estimates that 
efficiently combines diverse computational methods. We use the \texttt{emcee} sampler for 
best-fit estimates \footnote{For the \texttt{emcee} fitting, we ensure that the total number of 
samples in the chain exceeds at least 50 times the autocorrelation function (Nsamples / $\tau$ > 50) 
by using more than twice as many walkers as there are free parameters. The initial $20\%$ of the 
chain is then discarded as burn-in.} (Foreman Mackey 2013), where the $\chi^2$ statistic 
serves as a measure of the goodness of fit. The MCMC samples are visualized using \texttt{getdist} 
(Lewis 2019). The mean of the marginalized posterior distributions 
is estimated for all fits, along with probability contours for 68.3\% and 95.5\%. The  Pantheon+ 
data used in this work are publicy available on GitHub.
\footnote{See
Pantheon+: \mbox{\url{https://github.com/PantheonPlusSH0ES}}. 
}.

\bigskip

\noindent
During the fitting process, all model-dependent 
cosmological parameters are allowed to vary freely and are simultaneously fitted.
The results are mapped within the galactic coordinate system. For the main parameters of interest we adopt 
the same flat priors as those used by Brout et al. (2022), summarized in Table 2 . It 
should be noted that for Pantheon+ these are the only free parameters.

\begin{table}[htbp]
\centering
\begin{tabular}{lc}
\multicolumn{2}{l}{\textbf{Pantheon+}} \\
\hline
$\Omega_m$ & [0.1, 0.9] \\
$h$ & [0.55, 0.91] \\
$M$ & [-20.0, -18.0] \\
\hline
\end{tabular}
\caption{Flat priors for the cosmological parameters in each dataset.}
\end{table}

\bigskip

\noindent
The idea behind the RF method is to 
generate multiple randomly distributed directions, $(l_i,b_i)$, which are then used to select 
\mbox{SNe Ia} within specific regions, forming distinct subsamples in which cosmological 
parameters are fitted for each sky direction. A region is defined by the condition 
$\theta < \theta_\text{max}$, where $\theta$ is the screening angle, representing the angular 
separation between the random direction and the position of each SN Ia, with 
$\theta_\text{max} = 90^\circ$ explored in the present study.

\bigskip

\noindent
SNe Ia data located in respect to specific regions  $\theta$ $<$ $\theta_{max}$,
with  ${\it l}$ $\epsilon$ (0$^{\circ}$, 360$^{\circ}$) and ${\it b}$
$\epsilon$ (-90$^{\circ}$, 90$^{\circ}$)
being the longitude and latitude of the SN Ia
in the galactic coordinate system and 
$\bf{D}(l,b)$
the vector direction between the SN Ia position and the selected
$\theta$, form subsamples.
In order to describe
the degree of deviation from the cosmological principle, 
the anisotropic level  $\it AL$  is  described as
(see, for instance, Hu et al. 2024):

\begin{equation}
AL_{max} = \frac{h_{max} - h_{min}}{\sqrt{\sigma^{2}_{h_{max}} +
     \sigma^{2}_{h_{min}}}}
\end{equation}

\bigskip

\noindent
Here, $h$ is the dimensionless Hubble constant, $h = H_{0}/100$ km s$^{-1}$ Mpc$^{-1}$,  
$h_\text{min}$ and $h_\text{max}$ are the minimum and maximum $h$ values of the best
fitting results, with the corresponding 
$\sigma_{hmin}$ and $\sigma_{hmax}$ 1$\sigma$ error.
$AL_{max}$ is calculated as well with a set of the original data spread
evenly in the sky.  In this isotropic data set, $AL$  is also calulated.
In this way, a better idea of the significance of the results is obtained. 

\bigskip

\noindent
We apply the RF method to the Pantheon+ dataset, analyzing 2000 distinct sky regions and using the full 
sample with a $z>0.01$ cut. This analysis is performed twice: first, using only the 
Cepheid calibrators located within each region; and second, using all 42 Cepheids to calibrate all
regions. Performing both analyses allows us to separate underlying anisotropy signals from potential 
artifacts arising from variations in the availability or spatial distribution of calibrators,  
which are limited in number, examining whether the identified preferred directions are influenced 
by the calibration sample size and distribution. We report the statistical significance of the 
directions of maximum anisotropy obtained from both approaches. The uncertainties on the inferred 
directions are estimated from the standard deviation of the directions within $30^\circ$ of angular 
separation from each preferred direction with their $h$ values within $1\sigma$, which is chosen 
as a reasonable estimate to quantify local directional trends.

\bigskip

\noindent
For Pantheon+, the availability of the full covariance matrix enables a statistically robust 
evaluation of significance through a Cholesky Decomposition of the covariance matrix of the 
data vectors corresponding to each pair of directions to be compared. This technique enables 
the proper application of the RF method by accounting for correlations between overlapping 
\mbox{SNe Ia} in different sky regions. The RF method extends and generalizes the HC method, 
originally proposed by Schwarz \& Weinhorst (2007). The HC method
basically differs from the RF  in that the regions are hemispheres
covering 180$^{\circ}$  degrees in the sky. The RF method
is particularly well suited for the Pantheon+ 
dataset, where the full covariance matrix is available.
The Cholesky Decomposition is a widely 
used technique in numerical linear algebra (Golub et al. 1996;
Johnson et al. 2014; Chapra et al. 2017), 
particularly for Monte Carlo simulations and statistical modeling, where it is applied to simulate 
systems with multiple correlated variables (Asmussen 2007). The method requires 
that the covariance matrix be symmetric and positive-definite. By decomposing the covariance 
matrix into a product of a lower triangular matrix and its transpose, the decomposition allows 
the generation of correlated samples that preserve the underlying covariance structure. 
This process is mathematically represented as:

\begin{equation}
C=LL^T
\end{equation}

\noindent
where $L$ is a lower triangular matrix.  This factorization ensures that the covariance structure 
is retained, allowing for statistically robust inferences about systematic effects and uncertainties. 
Different applications of the Cholesky Decomposition can be found in the literature. So, for instance, 
Nikakhtar et al. (2018) use it to generate Monte Carlo realizations of correlated random walks,
whereas Yuan \& Eisenstein (2019) apply it to de-correlate covariance matrices in galaxy clustering 
analyses.

\bigskip

\noindent
We further assess the robustness of the anisotropy directions derived from the Pantheon+ dataset 
by simulating observational residuals in the distance modulus. 
Following the representative distance-modulus uncertainties 
(corresponding to an approximate dispersion of 
\(4\ \mathrm{km\ s^{-1}\ Mpc^{-1}}\) in \(H_0\)), we generate four simulated realizations of the 
Pantheon+ dataset in which each supernovae distance modulus is randomly 
perturbed by the  $\sigma_{int}$ of the SN Ia method.

\begin{equation}
H_{0.i\pm} = H_{0,i}(1 + \delta_{i}) = H_{0,i} + \delta H_{0,i} 
\end{equation}

\noindent
 The  set of simulations 
considered (restricted by computational constraints) can  identify  
instabilities: if significant directional variations are already observed under a small 
number of realizations, this should provide evidence that the inferred anisotropy directions 
are sensitive to observational uncertainties. This stability test is applied to the RF method and the HC method using local 
Cepheid calibrators within each sky region (as in the first approach described above). It is also applied
using the full set of 42 Cepheids.  For each  method,  
we identify the new directions of maximum and minimum \(h\). Shifts between the anisotropy directions in the 
original and simulated datasets will indicate whether the observed anisotropies are consistent with residual uncertainties.

\section{Results from the Pantheon+ dataset}

\begin{figure}[htbp]
\centering
\includegraphics[width=.46\textwidth]{"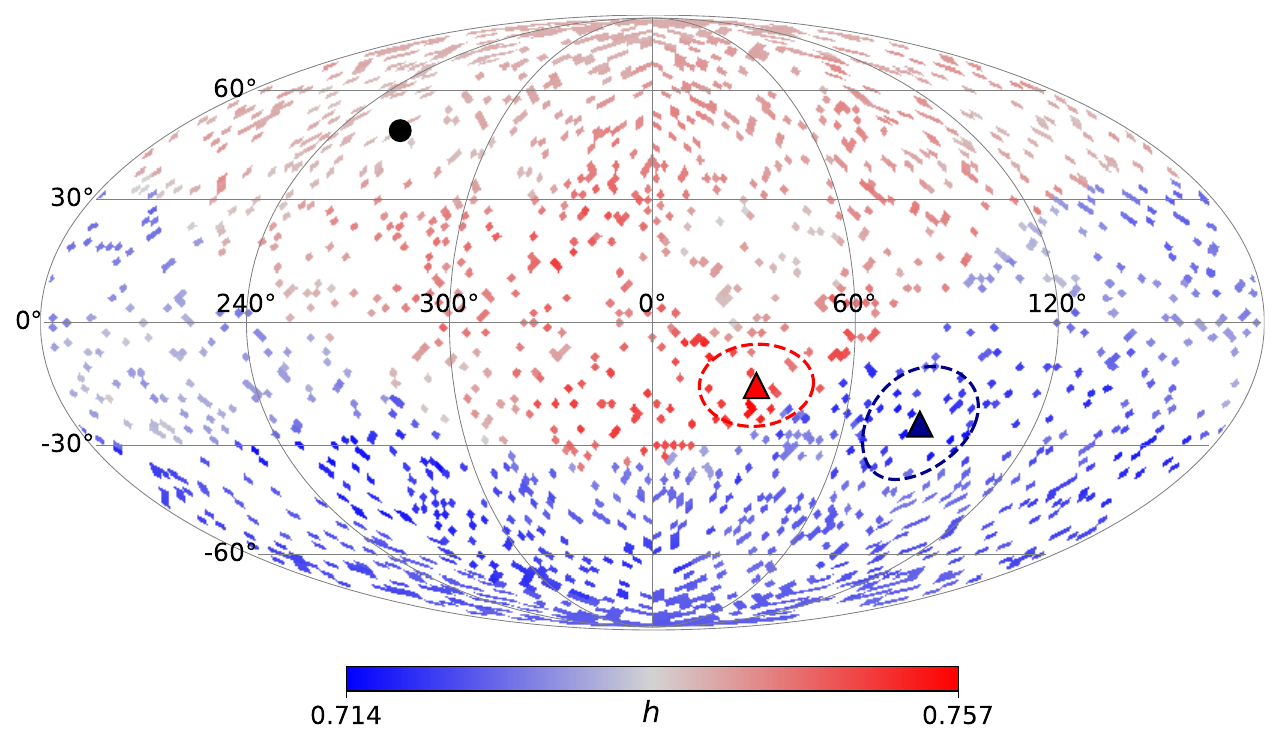"}
\quad
\includegraphics[width=.46\textwidth]{"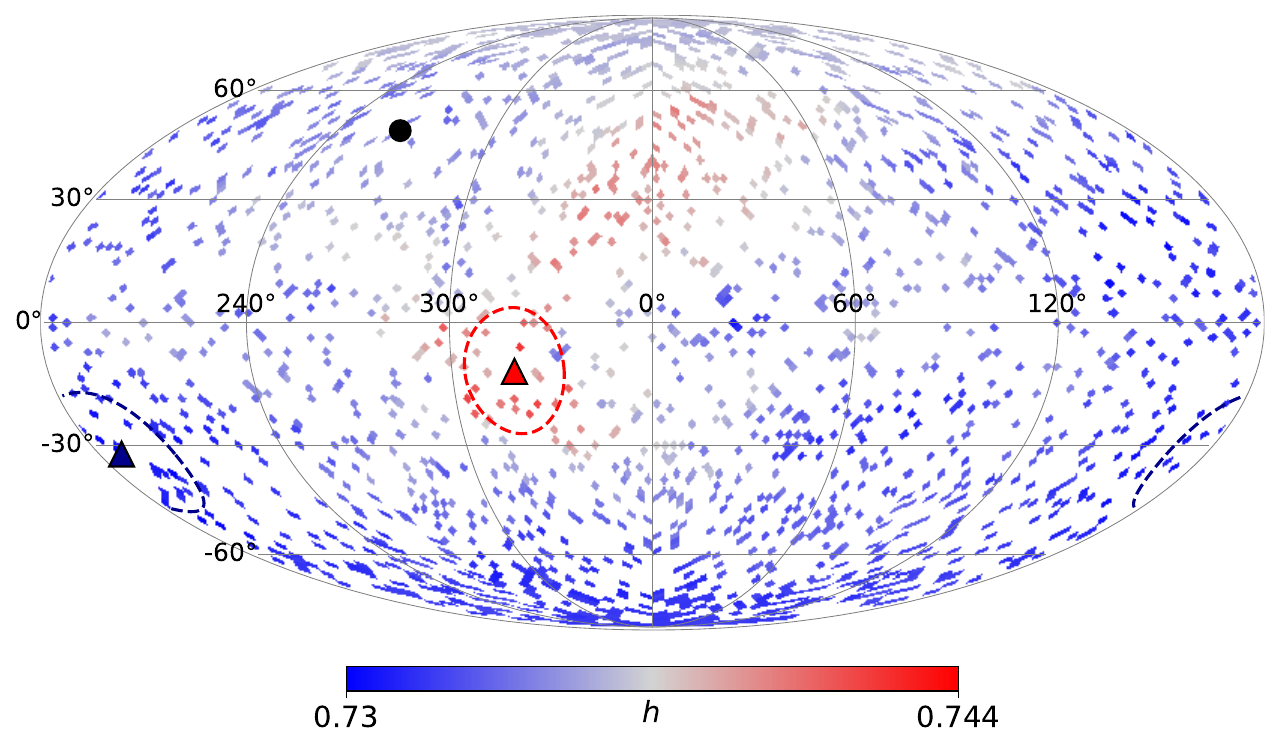"}
\caption{RF method applied to Pantheon+ dataset with only directional Cepheid calibrators (top), 
and with all 42 Cepheids calibrators in all directions (bottom). Panel show all-sky maps of $h$ 
for the 2000 directions investigated. Red and blue triangles mark the directions of maximum and 
minimum $h$, respectively. The black circle denotes the CMB dipole 
direction.
\label{Figure 5}}
\end{figure}

\noindent
The RF method using local Cepheids to calibrate each region yields an all-sky map 
with $h$ in the range $0.714$ to $0.757$, 
while using all 42 Cepheids for calibration produces a narrower range, from 
$0.730$ to $0.744$. For the local-Cepheid calibration, higher $h$ values are observed 
across a broad region covering much of the northern galactic hemisphere, coinciding with 
the CMB dipole direction, whereas lower values appear in the southern hemisphere. 
This hemispherical contrast is significantly attenuated when all 42 Cepheids are used 
for calibration, as observed in Figure 5. In particular, for the RF 
method using only local Cepheid calibrators, we find:

\[
h_\mathrm{max}^\text{Local} = 0.757 \pm 0.016, \qquad
h_\mathrm{min}^\text{Local} = 0.714 \pm 0.016,
\]

\noindent
at $(l,b)_{h_\mathrm{max}^\text{Local}} = (31.4 \pm 17.2, -15.2 \pm 10)$ 
and $(l,b)_{h_\mathrm{min}^\text{Local}} = (83.9 \pm 17.2, -24.6 \pm 14.1)$, respectively.

\bigskip

\noindent
When calibrating with all 42 Cepheids, the maximum and minimum values shift to: 

\[
h_\mathrm{max}^\text{All} =  0.744 \pm 0.011, \qquad
h_\mathrm{min}^\text{All} = 0.730 \pm 0.010,
\]

\noindent
at $(l,b)_{h_\mathrm{max}^\text{All}} = (318.7 \pm 14.9, -11.7 \pm 15.3)$ 
and $(l,b)_{h_\mathrm{min}^\text{All}} = (186.0 \pm 17.5, -32.2 \pm 15.2)$, respectively.

\bigskip

\noindent
Thus the preferred directions obtained with the two approaches differ. 
This suggests a dependence on the specific Cepheid sample considered. 
Since the second approach makes use of the full Cepheid dataset, it is expected to provide 
a more robust determination of the anisotropy directions and of the corresponding $h$ variation estimates.

\begin{figure}[htbp]
\centering
\includegraphics[width=0.4\textwidth]{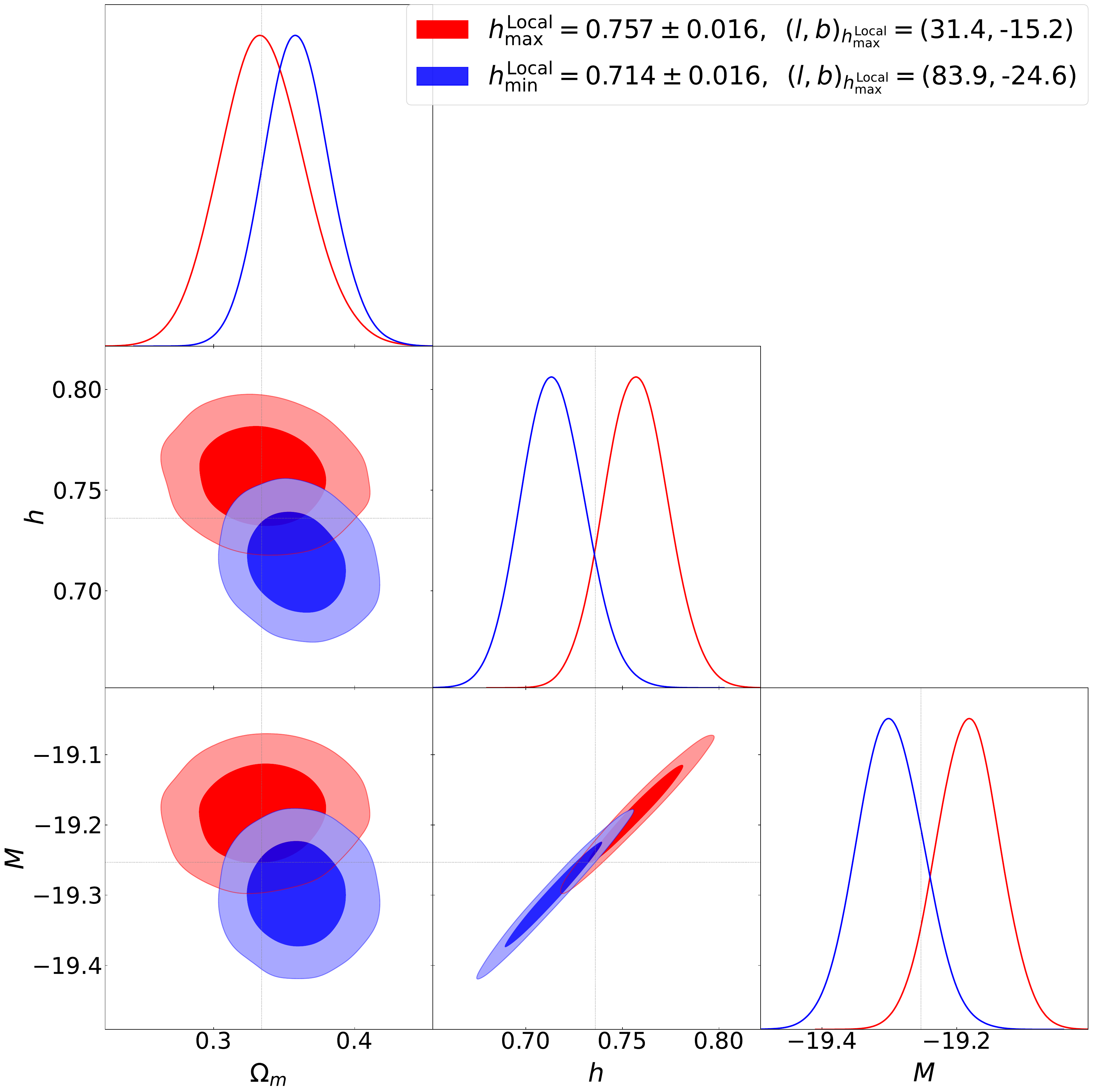}
\quad
\includegraphics[width=0.4\textwidth]{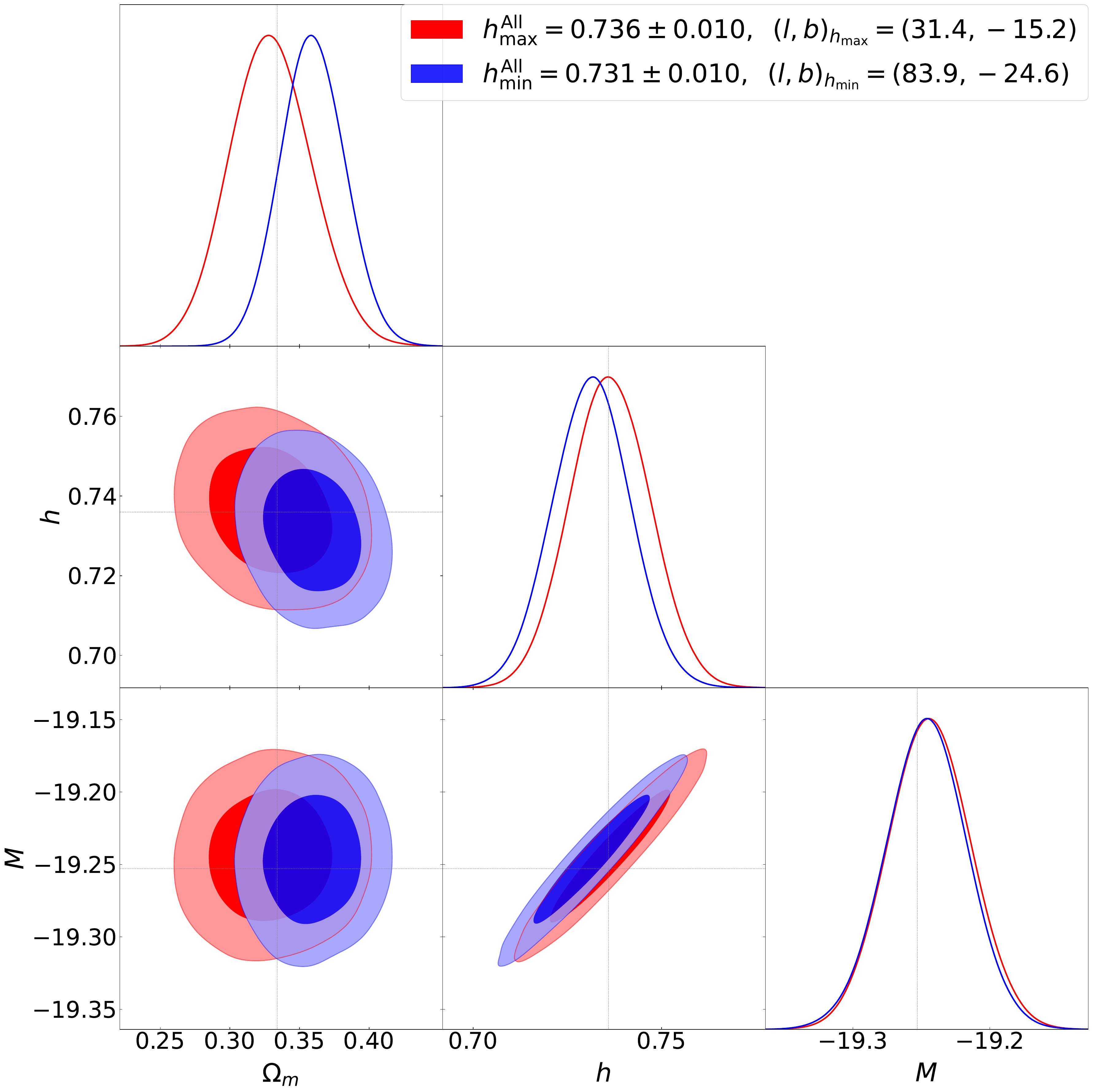}
\caption{Probability density contours for Pantheon+ with the RF method for the free parameters 
$\Omega_m$, $h$, and $M$, in the specific directions 
$(l,b)_{h_\mathrm{max}^\text{Local}}=(31.4 \pm 17.2,\,-15.2 \pm 10.0)$ and 
$(l,b)_{h_\mathrm{min}^\text{Local}}=(83.9 \pm 17.2,\,-24.6 \pm 14.1)$. 
Left: Using only local Cepheid calibrators. Right: Using all 42 Cepheid calibrators in both directions. 
Vertical and horizontal lines indicate the parameter values reported by Brout et al. (2022) from their 
analysis of the full Pantheon+ sample.}
\label{Figure 6}
\end{figure}

\begin{table*}[t]
\centering
\small
\begin{tabular}{lccc}
    \hline
    \textbf{RF method} & $\boldsymbol{h}\textbf{[}\boldsymbol{\Delta h}\textbf{]}$ &
    \textbf{Significance} & \textbf{A. direction $\boldsymbol{(l,b)}$} \\
    \hline
    \multicolumn{4}{c}{\textbf{Local Cepheids}} \\
    \hline
    Max + Min  & 0.732 $\pm$ 0.013 & - & - \\
    Max & 0.757 $\pm$ 0.016 [0.025] & $2.8\sigma$ in $h$ & $(31.4 \pm 17.2,\,-15.2 \pm 10.0)$ \\
    Min & 0.714 $\pm$ 0.016 [-0.018] & $1.8\sigma$ in $h$ & $(83.9 \pm 17.2,\,-24.6 \pm 14.1)$ \\
    \hline
    \multicolumn{4}{c}{\textbf{All 42 Cepheids}} \\
    \hline
    Max + Min  & 0.735 $\pm$ 0.010 & - & - \\
    Max & 0.744 $\pm$ 0.011 [0.009] & $<1\sigma$ in $h$ & $(318.7 \pm 14.9,\,-11.7 \pm 15.3)$ \\
    Min & 0.730 $\pm$ 0.010 [-0.005] & $<1\sigma$ in $h$ & $(186.0 \pm 17.5,\,-32.2 \pm 15.2)$ \\
    \hline
\end{tabular}
\caption{Hubble parameter estimates for Pantheon+ with the RF method. Top: using local Cepheids to 
  calibrate each region; bottom: using all 42 Cepheids.
This table reports the significance of $h_{\max}$ and $h_{\min}$ relative to the combined 
Max+Min sample, which denotes the reference dataset constructed by combining all SNe Ia belonging to 
the $h_{\max}$ and $h_{\min}$ regions, including those shared by both datasets. The cosmological parameters are fitted 
to this combined sample to obtain the reference value $h_{\max+\min}$ between the two anisotropy directions. Since the 
$h_{\max}$ and $h_{\min}$ measurements are not independent of the combined sample due to the large overlap in SNe Ia, we 
quantify the significance of the shifts by generating 100 independent realizations of the full dataset for each case 
(local and all Cepheids) using a Cholesky decomposition of the full covariance matrix. For each realization, we estimate 
$h_{\max+\min}$, $h_{\max}$, and $h_{\min}$, and compute the shifts $\Delta h = h_{\max} - h_{\max+\min}$ and 
$\Delta h = h_{\min} - h_{\max+\min}$. The standard deviation $\sigma(\Delta h)$ is then obtained from the distribution 
of simulated shifts, and the significance is defined as the ratio between the measured $\Delta h$ and $\sigma(\Delta h)$. 
For example, in the local Cepheid case for the maximum direction, we find $\Delta h = 0.025$ and $\sigma(\Delta h) \simeq 0.00886$, 
corresponding to a significance of $\sim 2.8\sigma$. This approach properly accounts for the correlations between subsamples and 
avoids an incorrect quadrature combination of uncertainties.
}
\end{table*}

\bigskip

\noindent
To evaluate the statistical significance of the maximum and minimum $h$ values for each case, we apply the 
Cholesky Decomposition methodology described in Section 3. 
Starting from the covariance matrix of the combined Max+Min sample, we generate 100 
realizations and perform the parameter inference for the combined, maximum, and minimum subsamples. 
From these simulations, we construct the distributions of the shifts with respect to the combined 
sample and use their standard deviations to quantify the statistical significance of the observed differences. 
Using local Cepheids, the combined Max+Min sample yields $h_{\mathrm{Max+Min}}^\text{Local} = 0.732 \pm 0.013$, 
with the statistical significance of the observed shifts being $2.8\sigma$ for the maximum direction 
and $1.8\sigma$ for the minimum. When calibrating with all 42 Cepheids, the combined sample 
gives $h_{\mathrm{Max+Min}}^\text{All} = 0.735 \pm 0.010$, with the corresponding statistical significance of 
the maximum and minimum directions being smaller than $1\sigma$ in both cases, as depicted in 
Table 2. One important finding is that the $2.8\sigma$ significance of
$h_{\rm max}^\text{Local}$ 
found using only the local Cepheid calibrators disappears when all 42 Cepheids are used at the 
same direction, with the value shifting from $0.757 \pm 0.016$ to $0.736 \pm 0.010$. Similarly, 
the $1.8\sigma$ significance of $h_{\rm min}^\text{Local}$ vanishes when evaluated with all Cepheid calibrators,
changing from $0.714 \pm 0.016$ to $0.731 \pm 0.010$. The corresponding 
corner plots for these two cases are shown in Figure 6.

\bigskip

\noindent
To assess the robustness of the inferred anisotropy directions, we introduce random perturbations 
$H_{0.i\pm} = H_{0,i}(1 + \delta_{i})$. We quote in Table 3 the results with the RF and HC
where the whole or the local SNe Ia Cepheids have been considered.

\begin{figure}[htbp]
\centering
\includegraphics[width=0.46\textwidth]{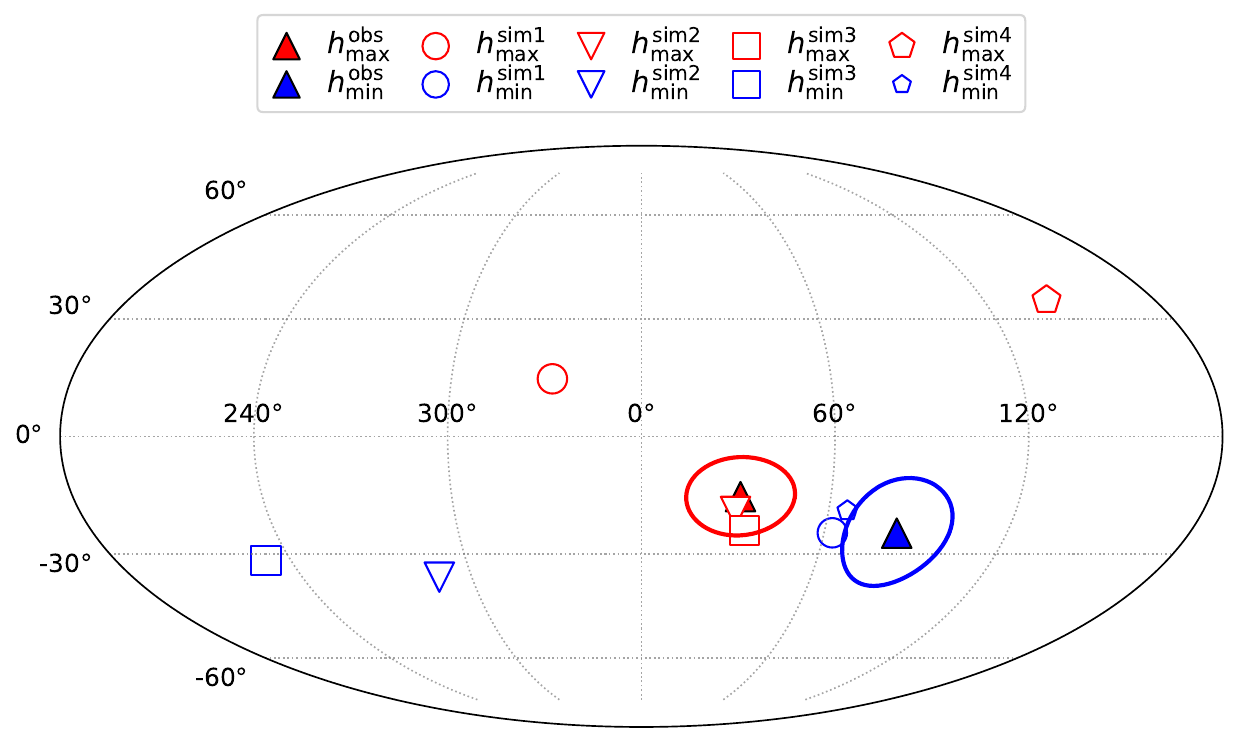}
\quad
\includegraphics[width=0.46\textwidth]{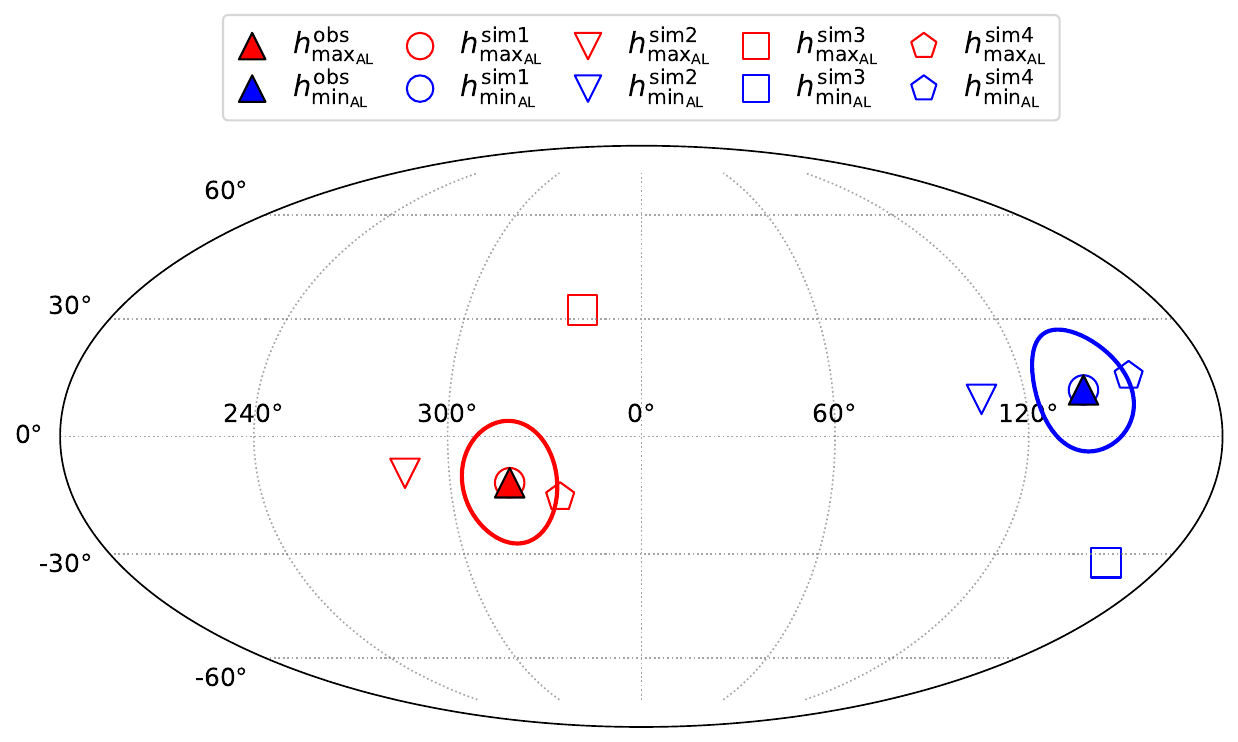}
\caption{Results of simulations including the intrinsic scatter of the
  SNe Ia method 
for Pantheon+. Top: RF method with local Cepheid calibrators; 
Bottom: HC method with all 42 Cepheids.}
\label{Figure 7}
\end{figure}

\begin{table*}[t]
  \centering
  \footnotesize

\caption{RF method with local Cepheids and HC method with all 42 Cepheids, four intrinsic-scatter 
simulations ($\sigma_{int} \simeq 0.11$ mag) for Pantheon+. Values of $h$ corresponding to the maximum anisotropy 
and their associated directions are listed for each realization.}

\begin{tabular}{cccccc}
\hline
\textbf{Method} & \textbf{N} & $h_{\rm max}$ & $(l,b)_{h_\text{max}}$ & $h_{\rm min}$ & $(l,b)_{h_\text{min}}$ \\
\hline
\multirow{4}{*}{\shortstack{RF \\ Local Cepheids}} 
& 1 & $0.786 \pm 0.013$ & $(291.5, 25.2)$ & $0.695 \pm 0.017$ & $(65.8, -27.2)$ \\
& 2 & $0.785 \pm 0.018$ & $(58.0, -6.5)$ & $0.713 \pm 0.013$ & $(219.7, -25.4)$ \\
& 3 & $0.792 \pm 0.018$ & $(36.3, -20.2)$ & $0.679 \pm 0.015$ & $(230.0, -33.4)$ \\
& 4 & $0.752 \pm 0.013$ & $(142.1, 35.1)$ & $0.676 \pm 0.016$ & $(66.0, -18.8)$ \\
\hline
\multirow{4}{*}{\shortstack{HC \\ All Cepheids}} 
& 1 & $0.763 \pm 0.011$ & $(318.7, -11.7)$ & $0.745 \pm 0.010$ & $(138.7, 11.7)$ \\
& 2 & $0.748 \pm 0.011$ & $(286.2, -9.3)$ & $0.735 \pm 0.010$ & $(106.2, 9.3)$ \\
& 3 & $0.751 \pm 0.011$ & $(339.8, 32.4)$ & $0.736 \pm 0.011$ & $(159.8, -32.4)$ \\
& 4 & $0.733 \pm 0.010$ & $(334.3, -15.3)$ & $0.720 \pm 0.010$ & $(154.3, 15.3)$ \\
\hline
\end{tabular}
\end{table*}

\noindent
They are just reflecting the error in the SNe Ia lightcurve method, and
therefore are smaller than Pantheon+ residual uncertainties, for both
the RF method with local Cepheid calibrators and the HC method using all 42 Cepheids.  
In Table 3 we present four realizations of the Pantheon+ dataset with the 
applied perturbations for each approach. It can be seen that both maximum and minimum $h$ values, as well as the corresponding 
preferred directions, differ in most cases from those obtained in the unperturbed case. 
In the case of the RF method, the changes appear more irregular, 
reflecting lower stability in the inferred anisotropy directions. In contrast, for the HC method the directions 
do not vary in a completely random manner, but the maximum anisotropy directions tend to shift beyond the estimated 
uncertainty region. This suggests that neither method yields consistent anisotropy directions. 
Figure 7 shows the original preferred directions along with those 
obtained from the four perturbed realizations of the Pantheon+ dataset for each case.

\section{The nature of the SNe Ia sets}

\noindent
Previoulsy to the use of Pantheon+, as mentioned in the
introduction, samples with smaller amounts of SNe Ia such as Union2, JLA,
Constitution sample, amongst others, where used. 

\bigskip

\noindent
Those studies suggest that any anisotropy is expected to be more 
prominent at low redshift, and in particular Colin et al. (2019) find a 
significant dipolar anisotropy. In this context, the CSP sample,
which is almost 
entirely composed of nearby SNe Ia, provides a useful dataset to explore potential 
anisotropy signals in the local Universe. The CSP dataset has notably fewer SNe Ia than 
Pantheon+ (342 SNe Ia) and the full covariance matrix is not available. 
This precludes the use of the RF method, but it remains testable with the HC method. 
A similar limitation in the number of SNe Ia was present in the JLA sample (Betoule et al. 2014), 
for which isotropy analyses using the HC method have also been performed
(Lin, Wang, Chang 2016; Deng \& Wei 2018). 

\bigskip

\noindent
The CSP 
(Hamuy et al. 2006) combines observations from CSP-I and CSP-II, two consecutive campaigns which ran from 2004 to 2015. The CSP-II campaign extended its observations deeper into the Hubble flow, with its NIR spectroscopy program and overall survey strategy detailed in Phillips et al. (2019) and Hsiao et al. (2019). This dataset offers a homogeneously calibrated sample of SNe Ia, with distance scales anchored through Cepheids, Tip of the Red Giant Branch (TRGB), and Surface Brightness Fluctuations (SBF) calibrators. For consistency with the calibration approach adopted in Pantheon+, we base our analysis on the \mbox{Cepheid-calibrated} CSP sample in the B band maximum magnitudes. This sample contains SNe Ia including 25 hosts with Cepheid distances for anchoring the absolute magnitude, covering the redshift range up to $z < 0.138$. The calibration values and distance scale information are taken from
Uddin et al. (2024), which presents the latest combined analysis of CSP-I and CSP-II data for H$_0$ determination.
When we look at the individual errors in H$_{0}$ from the CSP  SNe Ia
we find as well 8--9 km s$^{-1}$ Mpc$^{-1}$, as it happens in Pantheon+. 
The part coming from the method is typically of  4 km s$^{-1}$ Mpc$^{-1}$ coming from
the intrinsic error of the SN Ia lightcurve method.
However, the larger variations, of the order of 8--9 km s$^{-1}$ Mpc$^{-1}$, are
the final
budget for each H$_{0,i}$ determination (see, for instance,
Ruiz-Lapuente et al. 2025).

\begin{figure}[htbp]
\centering
\includegraphics[width=.40\textwidth]{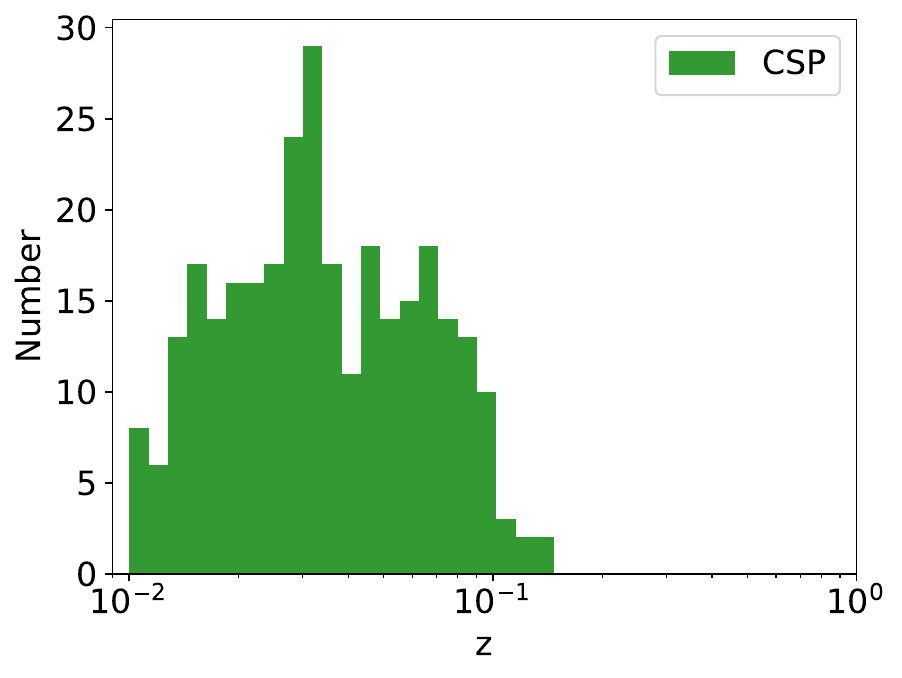}
\quad
\includegraphics[width=.46\textwidth]{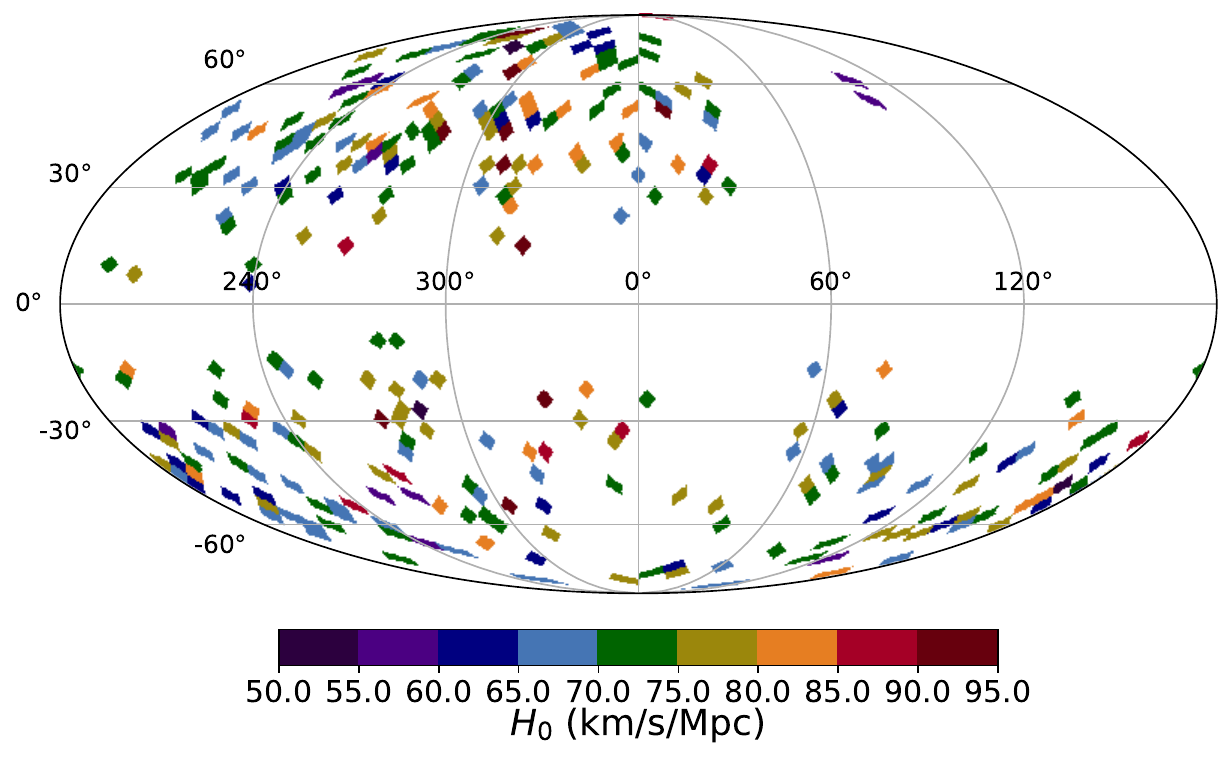}
\caption{Left: Redshift distribution of the CSP sample. Right: Distribution
  of SNe Ia in the sky, with the individual H$_{0,i}$ of each SN Ia. As in
  the case of Pantheon+, we find variations of 7-9 km s$^{-1}$ Mpc $^{-1}$
 from one H$_{0,i}$ to another one.}
\label{Figure 8}
\end{figure}

\bigskip

\noindent
We show the redshift distribution in Figure 8 and the values of H$_{0, i}$ for
each CSP SN Ia in the sky. 

\bigskip

\noindent
We apply the HC method to the CSP sample using all the  
25 Cepheid calibrators of the the CSP in each direction. The all-sky map of 
the Hubble parameter shows \( h \) to be in the range $0.701$ to $0.738$. From Figure 9 
it is observed that higher values of $h$ are obtained over a broad region encompassing 
the CMB dipole direction, similarly to the Pantheon+ dataset. The HC method, 
for the CSP sample, when accounting only for statistical uncertainties, 
yields a maximum anisotropy level of
 $2.7\text{AL}$ in $h$ between the opposite directions $(l,b)_{h_{\text{N}}} = 
 (354.5 \pm 12.2, 6.9 \pm 15.1)$ and $(l,b)_{h_{\text{S}}} = 
 (174.5 \pm 12.2, -6.9 \pm 15.1)$, 
 corresponding to the values
\[
h_N = 0.737 \pm 0.009, \qquad
h_S = 0.703 \pm 0.009.
\]

\begin{figure}[htbp]
\centering
\includegraphics[width=.46\textwidth]{"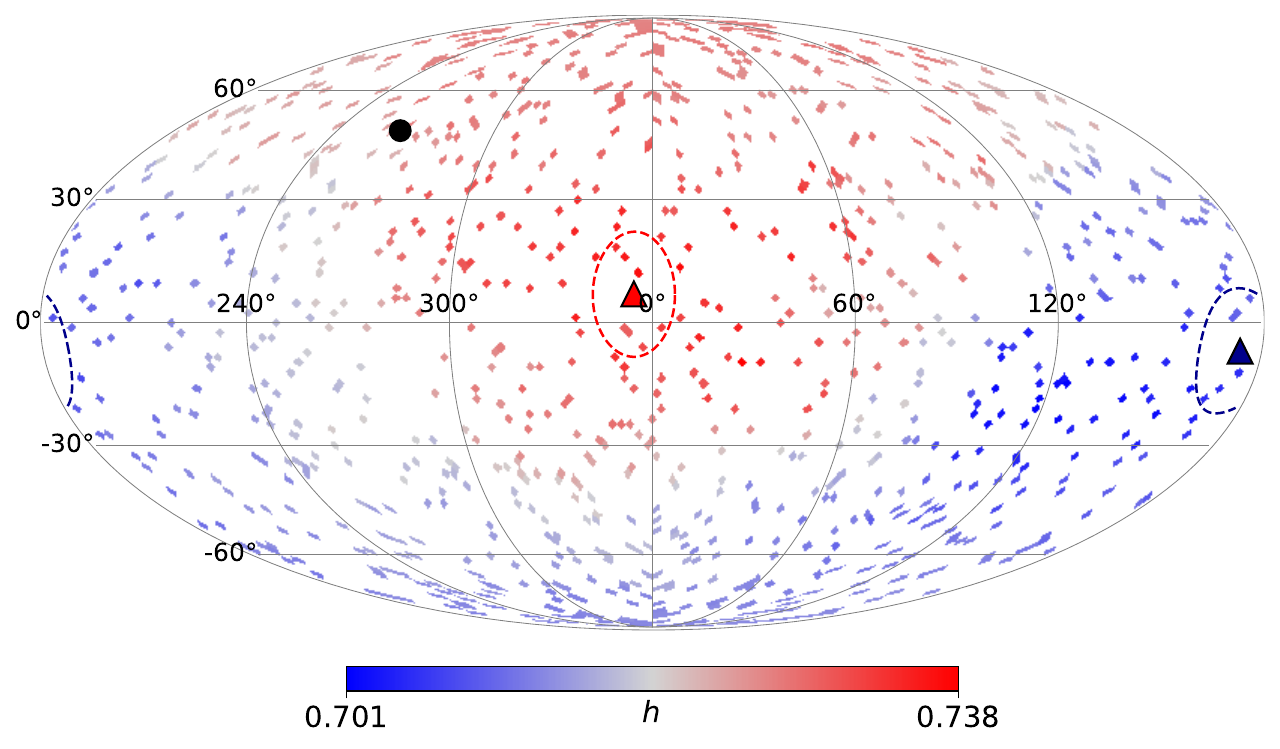"}
\caption{HC method applied to CSP dataset with all 25 Cepheids calibrators in all directions.
Panel show all-sky map of h for the 1000 directions investigated. Red and blue triangles mark the
antipodal directions of the maximum anisotropy level for h. The black circle denotes the CMB
dipole direction.}
\label{Figure 9}
\end{figure}

\bigskip

\noindent
To assess the confidence level of this result, we generate 30 statistically 
representative realizations of the dataset and compute the maximum anisotropy level for each realization, thus
obtaining a reference distribution for comparison. By following 
this approach, we find that the observed anisotropy level corresponds to a statistical 
significance of $2.5\sigma$. It should be emphasized that this result only accounts for 
statistical uncertainties.

\bigskip

\noindent
To account for systematic uncertainties, we adopt the approach from Uddin et al. (2024), where the systematic 
error is estimated as $\sigma_{h}^{\mathrm{sys}}=\frac{\sigma_{M}^{\mathrm{sys}}}{2.17}h$,
where the systematic uncertainty in the absolute magnitude is obtained by subtracting the statistical 
uncertainty from the total uncertainty in quadrature. We take the total uncertainty from Table~8 
of Uddin et al. (2024) being $\sigma_{M}^{\mathrm{stat+syst}} = 0.037$, corresponding to the Cepheid calibration 
in the $B$ band, and the statistical uncertainties are derived from our own fits. Following this procedure, we obtain 
a systematic uncertainty of $\sigma_{h}^{\mathrm{sys}} = 0.011$, for both maximum and minimum directions. 
Consequently, the significance of the observed anisotropy is reduced to the $1\sigma$ level, and the anisotropy is therefore 
not statistically compelling when the full uncertainty budget is taken into account.

\begin{table*}[t]
\centering
\footnotesize
\caption{Maximum hemispherical anisotropy in the Hubble parameter for the CSP sample.}
\begin{tabular}{lccccc}
    \hline
     & \multirow{2}{*}{\textbf{Value}} & \multirow{2}{*}{\textbf{AL}} & \multicolumn{2}{c}{\textbf{Significance}} \\
     &  &  & \textbf{Stat} & \textbf{Stat+Syst} \\
    \hline
    $h_{\text{AL}^{\text{max}}_N}$ & $0.737 \pm 0.009\,(\mathrm{Stat}) \pm 0.011\,(\mathrm{Syst})$ 
    & \multirow{2}{*}{$2.7\text{AL}$} 
    & \multirow{2}{*}{$2.5\sigma$} 
    & \multirow{2}{*}{$1.0\sigma$} \\
    
    $h_{\text{AL}^{\text{max}}_S}$ & $0.703 \pm 0.009\,(\mathrm{Stat}) \pm 0.011\,(\mathrm{Syst})$ 
    &  &  &  \\
    \hline
\end{tabular}
\end{table*}

\begin{table*}[t]
  \footnotesize
  \centering
  \caption{HC method with four intrinsic-scatter realizations of the CSP sample. Maximum and minimum $h$ values with sky coordinates $(l,b)$ are listed for each realization.}
  \begin{tabular}{c|c|cc|cc}
  \hline
  \textbf{Method} & \textbf{N} & $h_{\rm max}$ & $(l,b)_{h_\text{max}}$ & $h_{\rm min}$ & $(l,b)_{h_\text{min}}$ \\
  \hline
  \multirow{4}{*}{\shortstack{HC \\ All Cepheids}} 
  & 1 & $0.744 \pm 0.013$ & $(26.8, -10.7)$ & $0.722 \pm 0.009$ & $(206.8, 10.7)$ \\
  & 2 & $0.774 \pm 0.010$ & $(156.8, -21.9)$ & $0.714 \pm 0.010$ & $(336.8, 21.9)$ \\
  & 3 & $0.764 \pm 0.011$ & $(329.4, 16.0)$ & $0.748 \pm 0.010$ & $(149.4, -16.0)$ \\
  & 4 & $0.733 \pm 0.010$ & $(125.1, -14.0)$ & $0.720 \pm 0.010$ & $(305.1, 14.0)$ \\
  \hline
  \end{tabular}
\end{table*}

\begin{figure}[htbp]
\centering
\includegraphics[width=.40\textwidth]{"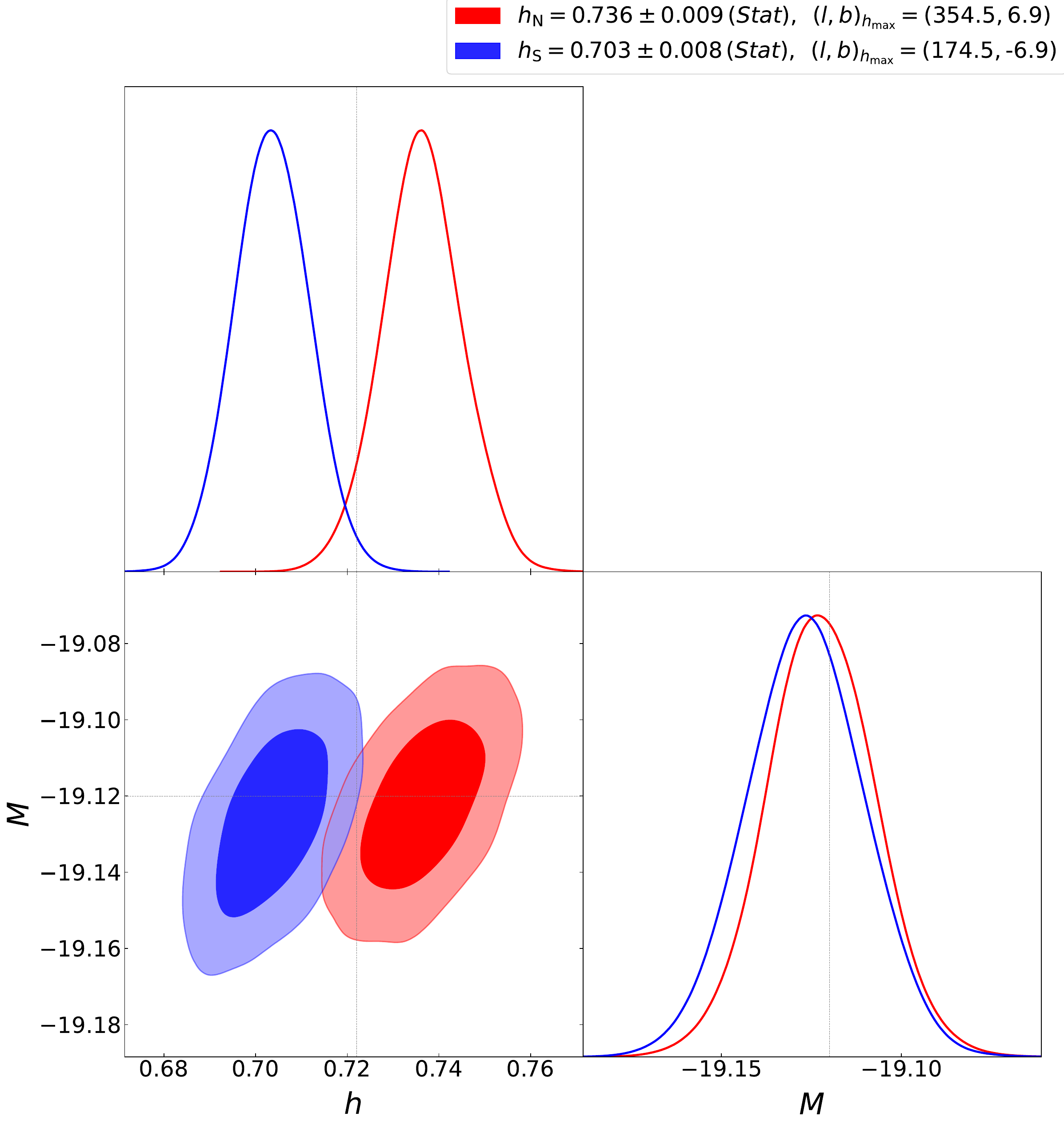"}
\quad
\includegraphics[width=.40\textwidth]{"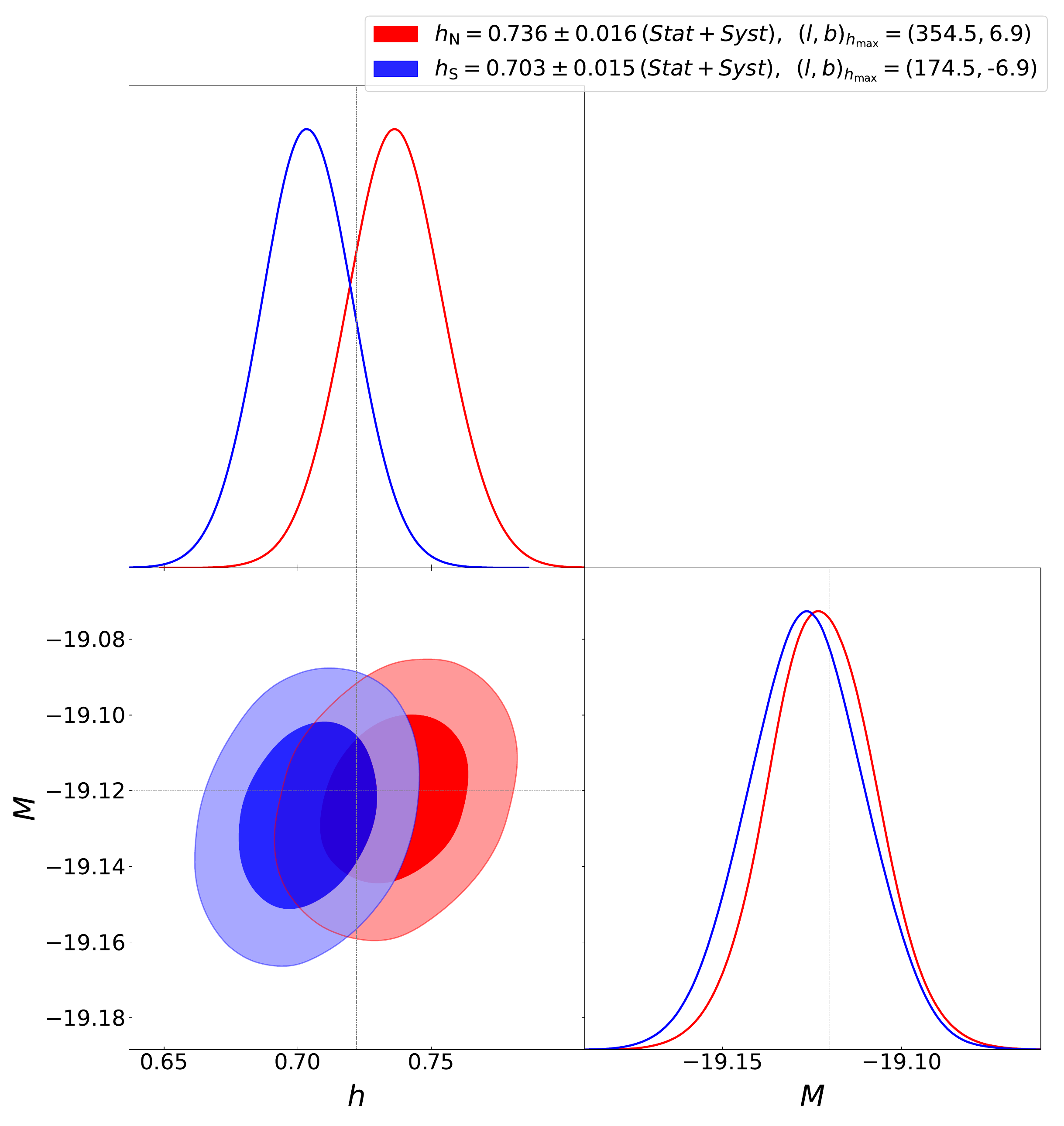"}
\caption{Probability density contours for CSP from the HC method for the free parameters 
$h$ and $M$, corresponding to the maximum anisotropy level in $h$ found along the opposite directions 
$(l,b)_{N} = (354.5 \pm 12.2,\, 6.9 \pm 15.1)$ and 
$(l,b)_{S} = (174.5 \pm 12.2,\,-6.9 \pm 15.1)$. 
Left: statistical uncertainties only. Right: statistical and systematic uncertainties. 
Vertical and horizontal lines indicate the parameter values reported by Uddin et al. (2024) 
from their analysis of the full CSP sample using Cepheid calibrators in the B band.
\label{Figure 10}}
\end{figure}

\bigskip

\noindent
We show in Figure 10 the corner plots of $h_{\text{AL}_{\text{max}}}$ 
and $h_{\text{AL}_{\text{min}}}$ for the CSP sample. The left panel shows 
results with statistical uncertainties only, while the right panel includes 
both statistical and systematic uncertainties. In each case, the 
plots display the joint probability of $h$ and $M$, marginalized 
over the remaining nuisance parameters.

\begin{figure}[htbp]
\centering
\includegraphics[width=.46\textwidth]{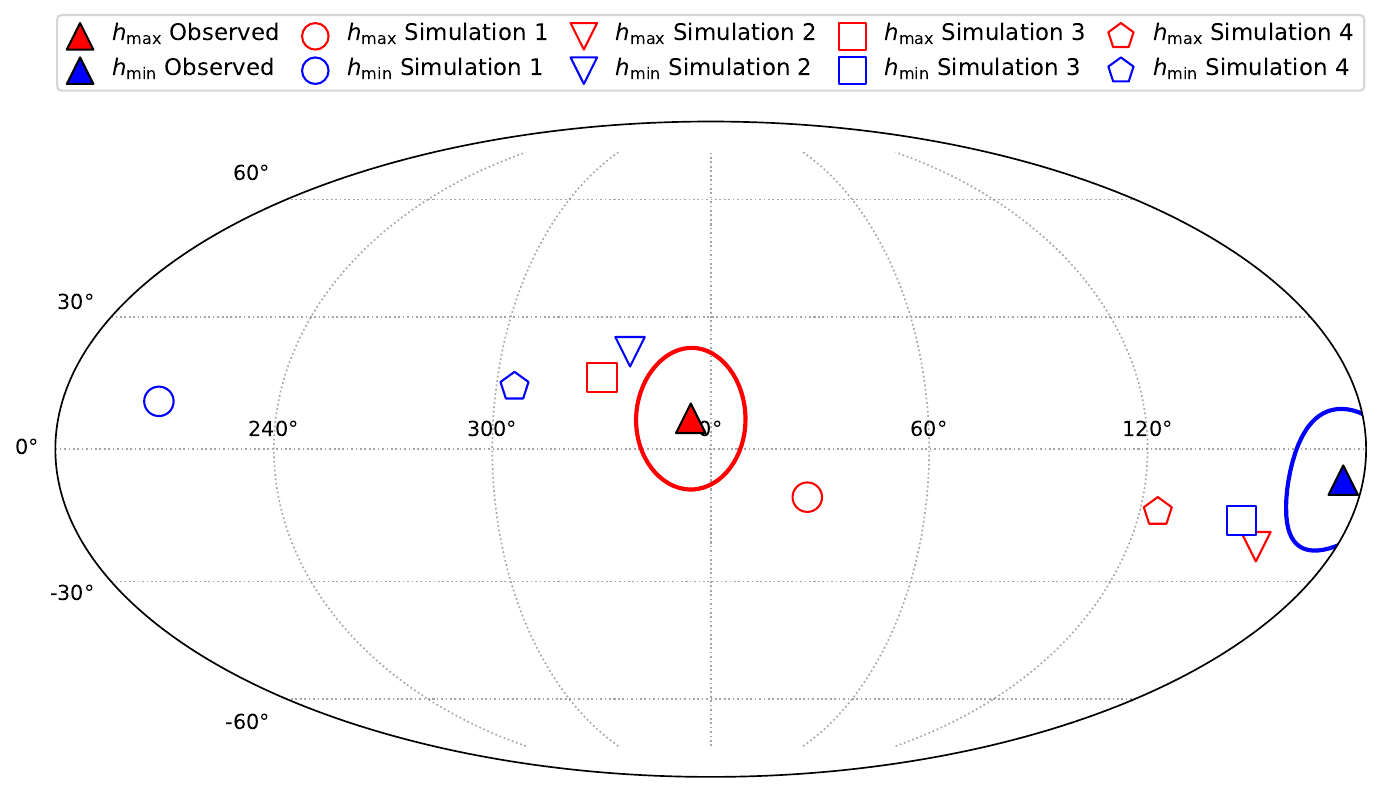}
\caption{Results of simulations including an intrinsic scatter of
 $\sigma_{int} = 0.11$ mag 
for CSP.}
\label{Figure 11}
\end{figure}

\bigskip

\noindent
Similarly to the analysis performed with the Pantheon+ sample, we investigate the stability of the inferred anisotropy 
directions using the CSP dataset by propagating the intrinsic scatter of SNe Ia through the full inference 
pipeline. In each Monte Carlo realization, every supernova is perturbed by a random intrinsic scatter term of 
amplitude $\pm 0.11$ mag, thatinducing variations in the inferred value of $h$.

\bigskip

\noindent
Table 5 summarizes our four representative realizations. We find significant variations in the anisotropy 
directions relative to the unperturbed case, indicating that the inferred axes are not robust against statistical and 
systematic fluctuations. The recovered directions for these realizations are shown in Fig. 11,
showing that the maximum anisotropy directions obtained with the HC method are weakly constrained. Despite this variability, 
the directions tend to align along an axis connecting the upper-left and lower-right regions of the galactic coordinate system. 
This apparent preference is consistent with the highly inhomogeneous sky distribution of CSP 
supernovae, which exhibit very limited coverage in the opposite (upper-right) region. Given that the HC method compares 
opposite hemispheres, this observational bias disfavors preferred axes intersecting poorly sampled areas. 
Therefore, the observed alignment is likely driven by the survey geometry rather than being a genuine cosmological signal.

\bigskip

\noindent
Table 5 summarizes four CSP realizations generated by applying 
the corresponding intrinsic scatter perturbations. Our results show significant variations 
in the values of $h$ and their associated directions when compared to the 
unperturbed case. This behavior suggests that the inferred anisotropy axes are not robust against 
statistical and systematic fluctuations.

\bigskip

\noindent
Figure 11 shows the directions obtained from the realizations 
including this intrinsic scatter. For clarity, only four representative realizations are displayed. 
The wide spread of these directions suggests that the recovered maximum anisotropy directions with the 
HC method are weakly constrained. 

\bigskip

\section{Discussion}

\noindent
Though, in most of the studies mentioned in the introduction, it has not been
fully tested the consistency in the direction of anisotropy of the SNe Ia,
we find interesting what is done 
 in Kazantzidis \& Perivolaropoulos (2020). They start with the
Pantheon sample which has 1048 SNe Ia that are quite non uniformly
distributed with a strong dependence of datapoint location in the
southern hemisphere in the longitude range [0$^{\circ}$, 180$^{\circ}$]. They
select a subsample of 375 SNe Ia and distribute them more isotropically in
the four quadrispheres (100 in the first three and 75 in the down right
quadrisphere). Using this reduced data set, which is more uniformly
distributed, they produce 100 simulated Pantheon subsamples using 1500
random directions to split it in two hemispheres. They identified
the maximum  magnitudes, finding opposite points in the maximum and
minimum directions. They show in their Figure 8 (left) the points of maximum
and minimum moving within the hemispheres with no preferent
direction. 

\bigskip

\noindent
The previous work had the limitation in the Pantheon data of not having the
calibration of distances. Pantheon+ has the advantage of being calibrated
withe the Cepheids from SH0ES. The magnitud $\mathcal{M}$ is a combination
that includes the absolute magnitude of SNe Ia M and the Hubble constant.

\bigskip

\noindent

\begin{equation}
\mathcal{M} \equiv M + 5\log_{10}\left[\frac{c/H_{0}}{1\,\mathrm{Mpc}}\right]
+ 25
\end{equation}

\bigskip

\noindent
The $\mathcal{M}$ is degenerate in M and  H$_{0}$. In the study of
evolution of M, $\mathcal{M}$ varying with z, could be as well H$_{0}$ varying with z. 
As the aim of this paper is exploring the evolution in z, the
creation of more uniform samples in the sky is positive for
the research.

\bigskip

\noindent
While the work by Kazantzidis \& Perivolaropoulos (2020), taking different
samples of 100 SNe Ia in every quadrisphere in the sky, implies that the
information on the position in the sky of this small samples will change,
the work by
Bengaly et al. (2024) fully redistributes an original Pantheon+ subsample
( SNe Ia from 0.01 $<$ z $<$ 0.1)  in an 
uniform distribution in the sky. They replace the original  SN coordinates
with random ones across the celestial sphere. 
They are interested in the
directional dependence of the parameter q$_{0}$. The
sample selected by them includes  697 SNe Ia from Pantheon+.
They make a Hemisphere Comparison,
but taking caps of 60$^{\circ}$ in the sky, and assume an M absolute magnitude of
the SNe Ia fixed, and H$_{0}$ fixed across directions in the sky.
With the original sample they obtain an anisotropic level of $\Delta$ q$_{0}$
$=$ 3.06. When they redistribute in the sky the SNe Ia, they obtain a
$\Delta$ q$_{0}$ $=$ 2.96. Thus, they conclude that there is no
statistically significant indication for a breakdown of the cosmic
isotropy  hypothesis  when testing q$_{0}$.

\bigskip

\noindent
In our case, 
substituting SNe Ia and placing them in a different sky position
would remove precisely variations in H$_{0}$ linked to existing
bulk flows created by inhomogeneities in mass distributions.
Thus, we do not redistribute them in the sky, for this test of directional
anisotropy, but we simply
perturb the  real sample with its real positions in the sky with 
a $\delta H_{0,i}$ from the light curve method. What would remain can then
be linked to important factors impacting on the calculation
of local H$_{0}$, including precisely the underlying velocity fields.

\bigskip

\noindent
We have seen that perturbing with  the individual  H$_{0,i}$,  the directional 
methods point to arbitrary anisotropy directions
in the sky.

\bigskip

\noindent
A number of works that use the HC method have referred to the effect
of having a sample  inhomogeneously distributed in the sky as a limitation
to judge on the anisotropy of the cosmological parameters. 
Deng and Wei (2018) provide an analysis using the HC method with the JLA
compilation, a sample of only 740 SNe Ia (Betoule et al. 2014). They do  as well the simulation
of having 5000 SNe Ia distributed with a given anysotropy (Pole--centralize or Equator centralized)
and find that the HC method is able to recover well the previously assigned anysotropy direction, whereas
another directional method named dipole fitting method fails. This exercise is
aimed at simulating what will happen with the Nancy Grace Roman Space Telescope SNe Ia,
at their time called WFIRST. However, they do not specify the error assigned in the determination
of the distance to SNe Ia with the Roman telescope (WFIRST in their paper).
Such error is expected to be  smaller than the present
one limited by the current use of the lightcurve SN Ia method. However,
the SNe Ia survey planned with Roman will not be an all sky survey. 
Another study is presented by Chang and Lin (2015), 
who analyze the Union2 dataset, a sample of 557 SNe Ia (Amanullah et al. 2010)
using the HC method.
These last authors find that the preferred directions 
inferred can differ significantly, even pointing in nearly opposite directions when considering 
subsets of the data. Through Monte Carlo simulations, they remark that the HC results are highly sensitive 
to the non-uniform angular distribution of SNe Ia, which can bias the inferred anisotropy signal.
A different approach to test isotropy is presented by Javanmardi et al. (2015), who analyse the Union2.1 compilation (Suzuki et al. 2012). Unlike the
standard HC technique, their method does not assume any specific form of anisotropy and is conceptually similar 
to the RF method. The sky is scanned by defining conical regions around different directions, and the SNe Ia 
within each cone are used to fit the cosmological parameters. Repeating this procedure over the full sky yields maps 
of $\Omega_\Lambda$, under the assumption of spatial flatness, together with a nuisance parameter absorbing $H_0$ and calibration effects. 
By varying the cone size ($\theta = 90^\circ,\, 60^\circ,\, 30^\circ$), they also test possible scale dependence 
of anisotropic signals. Their results are consistent with statistical fluctuations, and no robust evidence 
for anisotropy is found. While some directions show mild alignment with the CMB dipole direction, these features are not 
significant once sky coverage is taken into account, concluding that the Union2.1 sample is compatible with isotropy.
Zhao et al. (2019) applied the HC method to the original Pantheon compilation, covering the redshift range $0.01 < z < 2.3$. 
Since the original Pantheon sample is not calibrated with Cepheid distances, 
the analysis is performed using relative distance moduli. Using the HC method they report a preferred direction at 
$(l,b) \approx (123.1^\circ, 4.8^\circ)$, with a statistical significance of $2.1\sigma$ in $\Omega_m$. 
However, they show that this signal is predominantly driven by the low-z subset, indicating that it does 
not reflect a global anisotropic feature of the full dataset. They find a strong correlation between 
the inferred anisotropy map and the highly non-uniform angular distribution of SNe Ia in the Pantheon 
sample. Overall, their results suggest that any apparent 
anisotropy is highly sensitive to survey inhomogeneities. This conclusion is consistent with the 
limitations discussed in previous works.
These articles 
point to the fact that the identification of a cosmic preferred direction is data-dependent and should therefore
be interpreted with caution.

\bigskip

\noindent
Given the dependence on the size of the sample, now we will only refer to the largest SNe Ia
data set analysed for anisotropy: the Pantheon+ sample (Brout et al. 2022, Scolnic et al. 2022).
We have studied the factors that lead to conflicting confidence level in
a claim of anisotropy in Pantheon+. We can study 
 here  the results from 
 Hu et, al (2023), which use Pantheon+ and claim anisotropy at
 the 3.15--3.96$\sigma$ level. A major effect is to include the small
 amount from the
42 Cepheid calibrators that belong to each region of the fitting instead of all
Cepheids. This makes very small the amount of Cepheids for the calibration.
Using only local Cepheids per region in the sky  makes the distance estimates 
 highly sensitive to statistical fluctuations due to the
reduced subsample, 
and it could spuriously suggest anisotropy. If done in this way, the outcome
of the anisotropy in H$_{0}$ from the SNe Ia reveals
the anisotropy of the Cepheids.
Using the full calibrators sample provides a more robust
result (as presented here). It also
lowers the significance level to let the sample establish the absolute magnitude
M of the SNe Ia.  If M is fixed a  priori, with no allowance for any error,
one forces a  more significant result. In most papers M is let as a free
parameter, but it is not the case in Hu et al. (2023), for instance.
Just these changes (avoiding to introduce a false Cepheid anisotropy, and 
avoiding an a priori fixed M) bring the statistical significance
to $<$ 1$\sigma$, instead of 3$\sigma$.

\bigskip

\noindent
 McConville \& Colg\'ain (2023)
  treat $M$ as a free parameter. They impose, though,  a redshift 
cutoff of \(z < 0.7\). We find a comparable $h$ map distribution (noting that they use equatorial coordinates, 
whereas we adopt galactic coordinates). The region of $h_{max}$ coincides with the 
direction of the CMB dipole. 
 McConville \& Colg\'ain. (2023) report angular variations in $H_0$ 
with a significance of \(1.9\sigma\), interpreting their results as either a breakdown of isotropy or a 
statistical fluctuation among Cepheid-hosted supernovae. Even though we detect some level of anisotropy, reproducing their analysis 
with the RF method, our analysis indicates that this signal diminishes to $1\sigma$ when 
accounting for the full Cepheid calibration sample, that suggesting that the apparent 
anisotropy is definitely more likely due to limitations in the completeness of the Cepheid calibrating sample.

\bigskip

\noindent
Zhou, Dodelson \& Scolnic (2025)  have explored
as well perturbations in the map of individual H$_{0,i}$ 
of Pantheon+ SNe Ia
consistent with statistical noise and sample variance. In their research,
they consider the full sky $\delta$ map. The perturbation is larger than the one consider here, which was limited to  the intrinsic error of the SNe Ia
lightcurve method. Zhou, Dodelson \& Scolnic (2025) address as well the
uncertainty of the Hubble constant as measured in the CMB, $\delta_{CMB}$,
constructing residual maps of each sky patch
with  $\delta_{CMB}$ $=$ $h(p)/h_{Planck}-1$. They find  no hint of anisotropy
or correlation between early and late--Universe expansion with SNe Ia and CMB.

\bigskip

\noindent
Here we do not aim at studing the very early anisotropy of the Hubble
parameter, but the late--Universe one, and how the dozens of different applications of the region fitting or hemisphere comparison approaches miss
the point that
the direction marked by this method has no significant relevance in the context of the present
SNe Ia samples. 
 The  angular variation in the sky of 4 km s$^{-1}$ Mpc$^{-1}$ simply
 reflects half the scatter 
of values across different points in the galactic coordinate map.
It is important to keep 
in mind that this value is the
basic limitation in measuring $H_0$ from SNe Ia light curves. 

\bigskip

\noindent
When generating 
random realizations of the Pantheon+ and CSP datasets that include 
this intrinsic scatter, we find that the same RF and HC methods often yield significantly 
different preferred directions. Although the number of realizations is limited due to 
computational constraints, these results indicate that the directions identified by these 
methods are not stable in front of intrinsic scatter uncertainties. Therefore,
this result should be considered
when attempting to pinpoint a precise direction of anisotropy with such methods.

\bigskip

\noindent
The inferred anisotropy level is
fully consistent with isotropy at the current level of precision.
 The SNe Ia individual
H$_{0}$ should have smaller errors than the present ones. Then, it might be
possible to find at a reasonable confidence level any evidence of anisotropy.

\section{Conclusions}

\noindent
Our comprehensive Pantheon+ analysis, accounting for both statistical and systematic 
uncertainties, shows a weak departure from isotropy that vanishes when all Cepheid 
calibrators are considered in the maximum and minimum directions of $h$. Consequently, 
we find no compelling indication of a directional preference
in the Hubble parameter. We find, 
as it can be seen by analyzing the variation of $H_{0}$ along galactic 
coordinates, that there is no pattern privileging any direction of anisotropy. From one neighboring point 
to the next, $H_{0,i}$ is expected to  vary at least 4 km s$^{-1}$ Mpc$^{-1}$. We have found hints that the RF and HC 
methods can produce arbitrary directional anisotropy results. We performed the exercise of 
randomly distributing a point-to-point variation of $\pm$4 km s$^{-1}$ Mpc$^{-1}$
($\pm$8 km s$^{-1}$ Mpc$^{-1}$  is  the one observed in the 
CSP and Pantheon+ datasets, $\pm$4 km s$^{-1}$ Mpc$^{-1}$ corresponding to the intrinsic error of the lightcurve method, $\sigma_{int}$ $=$ 0.11 mag), and in each run of the RF and HC methods a different direction 
of anisotropy was obtained. This argues against the robustness of the directional results found 
by those methods when the scatter is large, as it is the case in our analysis.
Likewise, our statistical analysis using the CSP dataset, which probes the low-$z$ regime, 
shows a $\sim 2\sigma$ deviation. 
When the statistical and systematic errors are taken into account,
the anisotropy signal 
drops to the $1\sigma$ level. We highlight that with the RF and HC methods it is not possible 
to draw a definite conclusion about the presence or absence of anisotropy
in $H_{0}$ at present, with Pantheon+ and other SNe Ia samples. More robust 
constraints may require combining this approach with independent probes. 
Achieving robust constraints with SNe Ia will require lower SNe Ia
uncertainties, 
uniform sky coverage, and larger accurate SNe Ia samples that might be
soon available for rigorously testing 
the foundations of the standard cosmological model.

\begin{acknowledgements}

We acknowledge support from the High Performance Computing (HPC) resources of the DRAGO
supercomputer, affiliated with the Spanish National Research Council (CSIC)
PR-L and A.Q-E acknowledge support from grant PID2021-123528NB-I00, from the
the Spanish Ministry of Science and Innovation
(MICINN).
  
\end{acknowledgements}

\end{document}